%% file: Run_Chapter.tex
\documentclass[Alon2,ChapterTOCs,singlecolor,11pt]{Alon}
\usepackage{fixltx2e,fix-cm}
\usepackage[english]{babel}
\usepackage{epigrafica}
\usepackage[LGR,OT1]{fontenc}
\usepackage{lmodern}
\usepackage{amssymb}
\usepackage{amsmath}
\usepackage{graphicx}
\usepackage{subfigure}
\usepackage{makeidx}
\usepackage{multicol}

\usepackage{changepage}

\makeindex

\usepackage{version}

\title{Internet Traffic Analysis}
\author{Jeremy Kepner}

\include{cm-macro}
\begin{document}







\cleardoublepage
\setcounter{page}{7} 
\tableofcontents


\newcommand\blfootnote[1]{%
	\begingroup
	\renewcommand\thefootnote{}\footnote{#1}%
	\addtocounter{footnote}{-1}%
	\endgroup
}

\chapterauthor{Jeremy Kepner$^{1}$, Kenjiro Cho$^{2}$, KC Claffy$^{3}$, Vijay Gadepally$^{1}$, Sarah McGuire$^{1}$, Lauren Milechin$^{4}$, William Arcand$^{1}$, David Bestor$^{1}$, William Bergeron$^{1}$, Chansup Byun$^{1}$, Matthew Hubbell$^{1}$, Michael Houle$^{1}$, Michael Jones$^{1}$, Andrew Prout$^{1}$, Albert Reuther$^{1}$, Antonio Rosa$^{1}$, Siddharth Samsi$^{1}$, Charles Yee$^{1}$, Peter Michaleas$^{1}$}
{$^{1}$MIT Lincoln Laboratory, $^{2}$Research Laboratory, Internet Initiative Japan, Inc., $^{3}$UCSD Center for Applied Internet Data Analysis, $^{4}$MIT Dept. of Earth, Atmospheric and Planetary Sciences}

\chapter[New Phenomena in Large-Scale Internet Traffic]{New Phenomena in Large-Scale Internet Traffic}
\blfootnote{This material is based, in part, upon the work supported by the NSF under grants DMS-1312831, CCF-1533644, and CNS-1513283, DHS cooperative agreement FA8750-18-2-0049, and ASD(R\&E) under contract FA8702-15-D-0001.  Any opinions, findings, and conclusions or recommendations expressed in this material are those of the authors and do not necessarily reflect the views of the NSF, DHS, or ASD(R\&E).}

\chapterinitial{The}{Internet} is transforming our society, necessitating a quantitative understanding of Internet traffic.  Our team collects and curates the largest publicly available Internet traffic data sets. An analysis of 50 billion packets using 10,000 processors in the MIT SuperCloud reveals a new phenomenon: the importance of otherwise unseen leaf nodes and isolated links in Internet traffic. Our analysis further shows that a two-parameter modified Zipf--Mandelbrot distribution accurately describes a wide variety of source/destination statistics on moving sample windows ranging from 100{,}000 to 100{,}000{,}000 packets over collections that span years and continents. The measured model parameters distinguish different network streams, and the model leaf parameter strongly correlates with the fraction of the traffic in different underlying network topologies.

\section{Introduction}\label{intro}

Our civilization is now dependent on the Internet, necessitating a scientific understanding of this virtual universe \cite{hilbert2011world,li2013survey}.  The two largest efforts to capture, curate, and share Internet packet traffic data for scientific analysis are led by our team via the Widely Integrated Distributed Environment (WIDE) project \cite{cho2000tr} and the Center for Applied Internet Data Analysis (CAIDA) \cite{claffy1999internet}.  These data have been used for a wide variety of research projects, resulting in hundreds of peer-reviewed publications \cite{CAIDApubs}, ranging from characterizing the global state of Internet traffic, to specific studies on the prevalence of peer-to-peer file sharing traffic, to testing prototype software designed to stop the spread of Internet worms.

The stochastic network structure of Internet traffic is a core property of great interest to a wide range of Internet stakeholders \cite{li2013survey} and network scientists \cite{barabasi2016network}.  Of particular interest is the probability distribution $p(d)$, where $d$ is the degree (or count) of several network quantities, such as source packets, packets over a unique source-destination pair (or link),  and destination packets collected over specified time intervals. Among the earliest and most widely cited results of virtual Internet topology analysis has been the observation that $p(d) \propto 1/d^\alpha$ with a model exponent $1 < \alpha < 3$ for large values of $d$ \cite{barabasi1999emergence,albert1999internet,leskovec2005graphs} fit a range of network characteristics.

Many Internet models are based on the data obtained from crawling the network from a number of starting points \cite{olston2010web}.  These webcrawls naturally sample the supernodes of the network \cite{cao2009identifying}, and their resulting $p(d)$ are accurately fit at large values of $d$ by single-parameter power-law models. However, as we will show, for our streaming samples of the Internet there are other topologies that contribute significant traffic.  Characterizing a network by a single power-law exponent provides one view of Internet phenomena, but more accurate and complex models are required to understand the diverse topologies seen in streaming samples of the Internet.  Improving model accuracy while also increasing model complexity requires overcoming a number of challenges, including acquisition of larger, rigorously collected data sets \cite{soule2004identify,zhang2005estimating}; the enormous computational cost of processing large network traffic graphs \cite{lumsdaine2007challenges,bader2013graph,tune2013internet}; careful filtering, binning, and normalization of the data; and fitting of nonlinear models to the data.

\section{Methodology}\label{method}
This work aims to improve model accuracy through several techniques.  First, for over a decade we have scientifically collected and curated the largest publicly available Internet packet traffic data sets and this work analyzes the very largest collections in our corpora containing 49.6 billion packets (Table~\ref{tab:TrafficData}).  Second, utilizing recent innovations in interactive supercomputing \cite{Kepner2009,reuther2018interactive}, matrix-based graph theory \cite{kolda2009tensor,kepner2011graph}, and big data mathematics (Figure~\ref{fig:AssociativeArrays}) \cite{kepner2018mathematics}, we have developed a scalable Internet traffic processing pipeline that runs efficiently on more than 10{,}000 processors in the MIT SuperCloud \cite{gadepally2018hyperscaling}.  This pipeline allows us, for the first time, to process our largest traffic collections as network traffic graphs.  Third, since not all packets have both source and destination Internet Protocol version 4 (IPv4) addresses, the data have been filtered so that for any chosen time window all data sets have the same number of valid IPv4 packets, denoted $N_V$ (Figure~\ref{fig:ValidPackets} and Eq.~\ref{eq:Valid}). All computed probability distributions also use the same binary logarithmic binning to allow for consistent statistical comparison across data sets (Eq.~\ref{eq:Cumulative})\cite{clauset2009power,barabasi2016network}.  Fourth, to accurately model the data over the full range of $d$, we employ a modified Zipf--Mandelbrot distribution \cite{mandelbrot1953informational,montemurro2001beyond,saleh2006modeling}

\begin{equation}\label{eq:ZipfMandelbrot}
p(d;\alpha,\delta) \propto 1/(d + \delta)^\alpha
\end{equation}
The inclusion of a second model offset parameter $\delta$ allows the model to accurately fit small values of $d$, in particular $d=1$, which has the highest observed probability in these streaming data.  The modified Zipf--Mandelbrot model is a special case of the more general saturation/cutoff models used to model a variety of network phenomena (Eq.~\ref{eq:satcut}) \cite{clauset2009power,barabasi2016network}. Finally, nonlinear fitting techniques are used to achieve quality fits over the entire range of $d$ (Eq.~\ref{eq:NonLinFit}).

Throughout this chapter, we defer to the terminology of network science. Network operators use many similar terms with significant differences in meaning. We use network topology to refer to the graph-theoretic virtual topology of sources and destinations observed communicating and not the underlying physical topology of the Internet (Table~\ref{tab:Terms})

\begin{table}[htp]
	\caption{Network Terminology Used by Computer Network Operators and Network Scientists.  Throughout this Work, the Network Science Meanings are Employed}
	\begin{center}
		\begin{tabular}{p{0.75in}p{2.15in}p{2.15in}}
			{Term} & {Network Operations Meaning} & {Network Science Meaning} \\
			\hline
			Network &
			The \emph{physical} links, wires, routers, switches, and endpoints used to transmit data. &
			Any system that can be represented as a \emph{graph} of connections (links/edges) among entities (nodes/vertices). \\
			\hline
			Topology &
			The \emph{layout} of the physical network. &
			The specific \emph{geometries} of a graph and its sub-graphs. \\
			\hline
			Stream &
			The \emph{flow of data} over a specific physical communication link. &
			A time-ordered \emph{sequence of pairs} of entities (nodes/vertices) representing distinction in time connections (links/edges) between entities. \\
			\hline
		\end{tabular}
	\end{center}
	\label{tab:Terms}
\end{table}%

\subsection{MAWI and CAIDA Internet Traffic Collection}\label{mawicaida}
 For the analysis in following sections, the data utilized are summarized in Table~\ref{tab:TrafficData} with data from Tokyo coming from the MAWI Internet Traffic Collection and the data from Chicago from the CAIDA Internet Traffic Collection.

The Tokyo data sets are publicly available packet traces provided by the WIDE project (aka the MAWI traces). The WIDE project is a research consortium in Japan established in 1988 \cite{cho2000tr}. The members of the project include network engineers, researchers, university students, and industrial partners. The focus of WIDE is on the empirical study of the large-scale Internet. WIDE operates an Internet testbed both for commercial traffic and for conducting research experiments.  These data have enabled quantitative analysis of Internet traffic spanning years illustrating trends such as the emergence of residential usage, peer-to-peer networks, probe scanning, and botnets \cite{cho2006impact,borgnat2009seven,fontugne2017scaling}. The Tokyo data sets are publicly available packet traces provided by the WIDE project (aka the MAWI traces).  The traces are collected from a 1~Gbps academic backbone connection in Japan.  The 2015 and 2017 data sets are 48-hour-long traces captured during December 2--3, 2015, and April 12--13, 2017, in JST. The IP addresses appearing in the traces are anonymized using a prefix-preserving method \cite{fan2004prefix}.

The MAWI repository is an ongoing collection of Internet traffic traces, captured within the WIDE backbone network (AS2500) that connects Japanese universities and research institutes to the Internet. Each trace consists of captured packets observed from within WIDE and includes the packet headers of each packet along with the captured timestamp. Anonymized versions of the traces (with anonymized IP addresses and with transport layer payload removed) are made publicly available at http://mawi.wide.ad.jp/.

WIDE carries a variety of traffic including academic and commercial traffic. These data have enabled quantitative analysis of Internet traffic spanning years illustrating trends such as the emergence of residential usage, peer-to-peer networks, probe scanning, and botnets \cite{cho2006impact,cho2008observing,fontugne2017scaling}.  WIDE is mostly dominated by HTTP traffic, but is influenced by global anomalies.  For example, Code Red, Blaster, and Sasser are worms that disrupted Internet traffic \cite{allman2007brief}. Of these, Sasser (2005) impacted MAWI traffic the most, accounting for two-thirds of packets at its peak. Conversely, the ICMP traffic surge in 2003 and the SYN Flood in 2012 were more local in nature, each revealing attacks on targets within WIDE that lasted several months.

CAIDA collects several different data types at geographically and topologically diverse locations and makes these data available to the research community to the extent possible while preserving the privacy of individuals and organizations who donate data or network access \cite{claffy1999internet,claffy2000measuring}. CAIDA has (and had) monitoring locations in Internet service providers (ISPs) in the USA.   CAIDA's passive traces data set contains traces collected from high-speed monitors on a commercial backbone link. The data collection started in April 2008 and is ongoing. These data are useful for research on the characteristics of Internet traffic, including application breakdown (based on TCP/IP ports), security events, geographic and topological distribution, flow volume, and duration. For an overview of all traces, see the trace statistics page \cite{CAIDAstats}.

Collectively, our consortium has enabled the scientific analysis of Internet traffic, resulting in hundreds of peer-reviewed publications with over 30,000 citations \cite{CAIDApubs}.  These include early work on Internet threats such as the Code Red worm \cite{moore2002code} and Slammer worm \cite{moore2003inside} and how quarantining might mitigate threats \cite{moore2003internet}.  Subsequent work explored various techniques, such as dispersion, for measuring Internet capacity and bandwidth  \cite{dovrolis2001packet,dovrolis2004packet,prasad2003bandwidth}.  The next major area of research provided significant results on the dispersal of the Internet via the emergence of peer-to-peer networks \cite{karagiannis2004transport,karagiannis2004p2p}, edge devices \cite{kohno2005remote}, and corresponding denial-of-service attacks \cite{moore2006inferring}, which drove the need for new ways to categorize traffic \cite{kim2008internet}.  The incorporation of network science and statistical physics concepts into the analysis of the Internet produced new results on the hyperbolic geometry of complex networks \cite{krioukov2009curvature,krioukov2010hyperbolic} and sustaining the Internet with hyperbolic mapping \cite{boguna2009navigating,boguna2009navigability,boguna2010sustaining}. Likewise, a new understanding also emerged on the identification of influential spreaders in complex networks \cite{kitsak2010identification}, the relationship of popularity versus similarity in growing networks \cite{papadopoulos2012popularity},  and overall network cosmology \cite{krioukov2012network}. More recent work has developed new ideas for Internet classification \cite{dainotti2012issues} and future data centric architectures \cite{zhang2014named}.

\begin{table}[htp]
	\caption{Network Traffic Packet Data Sets from MAWI (Tokyo data sets) and CAIDA (Chicago data sets) Collected at Different Times and Durations over Two Years}
	\begin{center}
		\begin{tabular}{lcccc}
			{Location} & {Date} & {Duration} & {Bandwidth} & {Packets} \\
			\hline
			Tokyo     & 2015 Dec 02 & 2 days &   $10^9$~ bits/second &   $17.0{\times}10^9$ \\ 
			Tokyo     & 2017 Apr 12 & 2 days &   $10^9$~ bits/second &   $16.8{\times}10^9$ \\ 
			Chicago A & 2016 Jan 21 & 1 hour & $10^{10}$ bits/second & ~~$2.0{\times}10^9$  \\ 
			Chicago A & 2016 Feb 18 & 1 hour & $10^{10}$ bits/second & ~~$2.0{\times}10^9$  \\ 
			Chicago A & 2016 Mar 17 & 1 hour & $10^{10}$ bits/second & ~~$1.8{\times}10^9$  \\ 
			Chicago A & 2016 Apr 06 & 1 hour & $10^{10}$ bits/second & ~~$1.8{\times}10^9$  \\ 
			Chicago B & 2016 Jan 21 & 1 hour & $10^{10}$ bits/second & ~~$2.3{\times}10^9$  \\ 
			Chicago B & 2016 Feb 18 & 1 hour & $10^{10}$ bits/second & ~~$1.7{\times}10^9$  \\ 
			Chicago B & 2016 Mar 17 & 1 hour & $10^{10}$ bits/second & ~~$2.0{\times}10^9$  \\ 
			Chicago B & 2016 Apr 06 & 1 hour & $10^{10}$ bits/second & ~~$2.1{\times}10^9$  \\ 
			\hline
		\end{tabular}
	\end{center}
	\label{tab:TrafficData}
\end{table}%

\subsection{Network Quantities from Matrices}\label{networkquantities}
In our analysis, the network traffic packet data are reduced to origin--destination traffic matrices. These matrices can be used to compute a wide range of network statistics useful in the analysis, monitoring, and control of the Internet. Such an analysis includes the temporal fluctuations of the supernodes \cite{soule2004identify} and inferring the presence of unobserved traffic \cite{zhang2005estimating,bharti2010inferring}.

 To create the matrices, at a given time $t$, $N_V$ consecutive valid packets are aggregated from the traffic into a sparse matrix ${\bf A}_t$, where ${\bf A}_t(i,j)$ is the number of valid packets between the source $i$ and destination $j$ \cite{mucha2010community}. The sum of all the entries in ${\bf A}_t$ is equal to $N_V$
\begin{equation}\label{eq:Valid}
\sum_{i,j} {\bf A}_t(i,j) = N_V
\end{equation}
All the network quantities depicted in Figure~\ref{fig:NetworkDistribution}a can be readily computed from ${\bf A}_t$ as specified in Tables~\ref{tab:Aggregates} and ~\ref{tab:Filters}, including the number of unique sources and destinations, along with many other network statistics \cite{soule2004identify,zhang2005estimating,tune2013internet}.

\begin{table}
	\caption{Aggregate Network Properties}
		\begin{center}
		\begin{tabular}{p{1.5in}p{1.5in}p{1.0in}}
			\hline
			{Aggregate Property} & {Summation Notation} & {Matrix Notation} \\
			\hline
			Valid packets $N_V$ & $\sum_i ~ \sum_j ~ {\bf A}_t(i,j)$ & $~{\bf 1}^{\sf T} {\bf A}_t {\bf 1}$ \\
			Unique links & $\sum_i ~ \sum_j |{\bf A}_t(i,j)|_0$  & ${\bf 1}^{\sf T}|{\bf A}_t|_0 {\bf 1}$ \\
			Unique sources & $\sum_i |\sum_j ~ {\bf A}_t(i,j)|_0$  & ${\bf 1}^{\sf T}|{\bf A}_t {\bf 1}|_0$ \\
			Unique destinations & $\sum_j |\sum_i ~ {\bf A}_t(i,j)|_0$ & $|{\bf 1}^{\sf T} {\bf A}_t|_0 {\bf 1}$ \\
			\hline
		\end{tabular}{Formulas for computing aggregates from a sparse network image ${\bf A}_t$ at time $t$ in both summation and matrix notations. ${\bf 1}$ is a column vector of all 1's, $^{\sf T}$  is the transpose operation, and $|~|_0$ is the zero-norm that sets each nonzero value of its argument to 1 \cite{karvanen2003measuring}.}
	\end{center}
	\label{tab:Aggregates}
\end{table}%
\begin{table}[h]
	\caption{Neural Network Image Convolution Filters}
		\begin{center}
		\begin{tabular}{p{2.0in}p{1.5in}p{1.0in}}
			\hline
			{Network Quantity} & {Summation Notation} & {Matrix Notation} \\
			\hline
			Source packets from $i$ & $\sum_j ~ {\bf A}_t(i,j)$ & ~~~$~{\bf A}_t ~~ {\bf 1}$ \\
			Source fan-out from $i$ & $\sum_j |{\bf A}_t(i,j)|_0$  & ~~~$|{\bf A}_t|_0 {\bf 1}$ \\
			Link packets from $i$ to $j$ & $~~~~~~{\bf A}_t(i,j)$ & ~~~$~{\bf A}_t$ \\
			Destination fan-in to $j$ & $\sum_i |{\bf A}_t(i,j)|_0$ & ${\bf 1}^{\sf T}~{\bf A}_t$ \\
			Destination packets to $j$ & $\sum_i ~ {\bf A}_t(i,j)$ & ${\bf 1}^{\sf T}|{\bf A}_t|_0$ \\
			\hline
		\end{tabular}{Different network quantities are extracted from a sparse traffic image ${\bf A}_t$ at time $t$ via convolution with different filters.  Formulas for the filters are given in both summation and matrix notations. ${\bf 1}$ is a column vector of all 1's, $^{\sf T}$  is the transpose operation, and $|~|_0$ is the zero-norm that sets each nonzero value of its argument to 1 \cite{karvanen2003measuring}.}
	\end{center}
	\label{tab:Filters}
\end{table}%

Figure~\ref{fig:NetworkTopology}a depicts the major topological structures in the network traffic.  Isolated links are sources and destinations that each have only one connection (Table~\ref{tab:Isolatedlinks}).  The first, second, third, $\ldots$ supernodes are the source or destination with the first, second, third, $\ldots$ most links (Table~\ref{tab:Supernodes}).  The core of a network can be defined in a variety of ways \cite{schaeffer2007graph,benson2016higher}.  In this work, the network core conveys the concept of a collection of sources and destinations that are not isolated and are multiply connected.  The core is defined as the collection of sources and destinations in which every source and destination has more than one connection.  The core, as computed here, does not include the first five supernodes although only the first supernode is significant, and whether or not the other supernodes are included has minimal impact on the core in these data.  The core leaves are sources and destinations that have only one connection to a core source or destination (Tables~\ref{tab:Core} and \ref{tab:Coreleaves}).

\begin{table}[h]
	\caption{Properties of Isolated Links}
		\begin{center}
		\begin{tabular}{p{2.75in}p{1.25in}}
			\hline
			{Network Quantity} & {Matrix Notation} \\
			\hline
			Isolated links & ${\bf A}_t(i_1,j_1)$\\
			Number of isolated link sources & ${\bf 1}^{\sf T}|{\bf A}_t(i_1,j_1) {\bf 1}|_0$\\
		    Number of packets traversing isolated links &  ${\bf 1}^{\sf T}{\bf A}_t(i_1,j_1){\bf 1}$\\
			Number of unique isolated links & ${\bf 1}^{\sf T}|{\bf A}_t(i_1,j_1)|_0 {\bf 1}$ \\
			Number of isolated link destinations & $|{\bf 1}^{\sf T} {\bf A}_t(i_1,j_1)|_0 {\bf 1}$ \\
			\hline
		\end{tabular}{Different characteristics related to isolated links are extracted from a sparse traffic image ${\bf A}_t$ at time $t$.  Formulas are in matrix notation. The set of sources that send to only one destination are $ i_1 = {\rm arg}({\bf d}_{\rm out} = 1)$, and the set of destinations that receive from only one destination are $j_1 = {\rm arg}({\bf d}_{\rm in} = 1)$.}
	\end{center}
	\label{tab:Isolatedlinks}
\end{table}%

\begin{table}[h]
	\caption{Properties of Supernodes}
		\begin{center}
		\begin{tabular}{p{2.75in}p{2.25in}}
			\hline
			{\bf Network} & {\bf ~Matrix} \\
			{\bf Quantity} & {\bf Notation} \\
			\hline
			Supernode source leaves & ${\bf A}_t(i_1,k_{\rm max})$\\
			Supernode destination leaves & ${\bf A}_t(k_{\rm max},j_1)$\\
			Number of supernode leaf sources &  ${\bf 1}^{\sf T}|{\bf A}_t(i_1,k_{\rm max}) {\bf 1}|_0$\\
			Number of packets traversing supernode leaves & $ {\bf 1}^{\sf T}{\bf A}_t(i_1,k_{\rm max}) + {\bf A}_t(k_{\rm max},j_1){\bf 1}$ \\
			Number of unique supernode leaf links & $ {\bf 1}^{\sf T}|{\bf A}_t(i_1,k_{\rm max})|_0 + |{\bf A}_t(k_{\rm max},j_1)|_0 {\bf 1}$ \\
			Number of supernode leaf destinations & $|{\bf 1}^{\sf T} {\bf A}_t(k_{\rm max},j_1)|_0 {\bf 1}$\\
			\hline
		\end{tabular}{Different characteristics related to supernodes are extracted from a sparse traffic image ${\bf A}_t$ at time $t$.  Formulas are in matrix notation. The identity of the first supernode is given by $k_{\rm max} = {\rm argmax}({\bf d}_{\rm out} + {\bf d}_{\rm in})$. The leaves of a supernode are those sources and destinations whose only connection is to the supernode.}
	\end{center}
	\label{tab:Supernodes}
\end{table}%

\begin{table}[h]
	\caption{Properties of Network Core}
		\begin{center}
		\begin{tabular}{p{2.75in}p{2.25in}}
			\hline
			{\bf Network} & {\bf ~Matrix} \\
			{\bf Quantity} & {\bf Notation} \\
			\hline
			Core links & ${\bf A}_t(i_{\rm core},j_{\rm core})$\\
		    Number of core sources & ${\bf 1}^{\sf T}|{\bf A}_t(i_{\rm core},j_{\rm core}) {\bf 1}|_0$\\
		    Number of core packets & $ {\bf 1}^{\sf T}{\bf A}_t(i_{\rm core},j_{\rm core}){\bf 1}$\\
		    Number of unique core links & ${\bf 1}^{\sf T}|{\bf A}_t(i_{\rm core},j_{\rm core})|_0 {\bf 1}$\\
		    Number of core destinations & $|{\bf 1}^{\sf T} {\bf A}_t(i_{\rm core},j_{\rm core})|_0 {\bf 1}$\\
			\hline
		\end{tabular}{Different characteristics related to the core are extracted from a sparse traffic image ${\bf A}_t$ at time $t$.  Formulas are in matrix notation. The set of sources that send to more than one destination, excluding the supernode(s), is $i_{\rm core} = {\rm arg}(1 < {\bf d}_{\rm out} < {\bf d}_{\rm out}(k_{\rm max}))$.  The set of destinations that receive from more than one source, excluding the supernode(s), is $ j_{\rm core} = {\rm arg}(1 < {\bf d}_{\rm in} < {\bf d}_{\rm in}(k_{\rm max}))$.}
	\end{center}
	\label{tab:Core}
\end{table}%

\begin{table}[h]
	\caption{Properties of Core Leaves}
		\begin{center}
		\begin{tabular}{p{2.75in}p{2.25in}}
			\hline
			{\bf Network} & {\bf ~Matrix} \\
			{\bf Quantity} & {\bf Notation} \\
			\hline
			Core source leaves & ${\bf A}_t(i_1,k_{\rm core})$\\
		    Core destination leaves & ${\bf A}_t(k_{\rm core},j_1)$\\
		    Number of core leaf sources & ${\bf 1}^{\sf T}|{\bf A}_t(i_1,k_{\rm core}) {\bf 1}|_0$\\
		    Number of core leaf packets & ${\bf 1}^{\sf T}{\bf A}_t(i_1,k_{\rm core}) + {\bf A}_t(k_{\rm core},j_1){\bf 1}$\\
		    Number of unique core leaf links & ${\bf 1}^{\sf T}|{\bf A}_t(i_1,k_{\rm core})|_0 + |{\bf A}_t(k_{\rm core},j_1)|_0 {\bf 1}$\\
		    Number of core leaf destination & $ |{\bf 1}^{\sf T} {\bf A}_t(k_{\rm core},j_1)|_0 {\bf 1}$\\
			\hline
		\end{tabular}{Different characteristics related to the core leaves are extracted from a sparse traffic image ${\bf A}_t$ at time $t$.  Formulas are in matrix notation. The core leaves are sources and destinations that have one connection to a core source or destination.}
	\end{center}
	\label{tab:Coreleaves}
\end{table}%

\begin{figure}
	\includegraphics[width=\columnwidth]{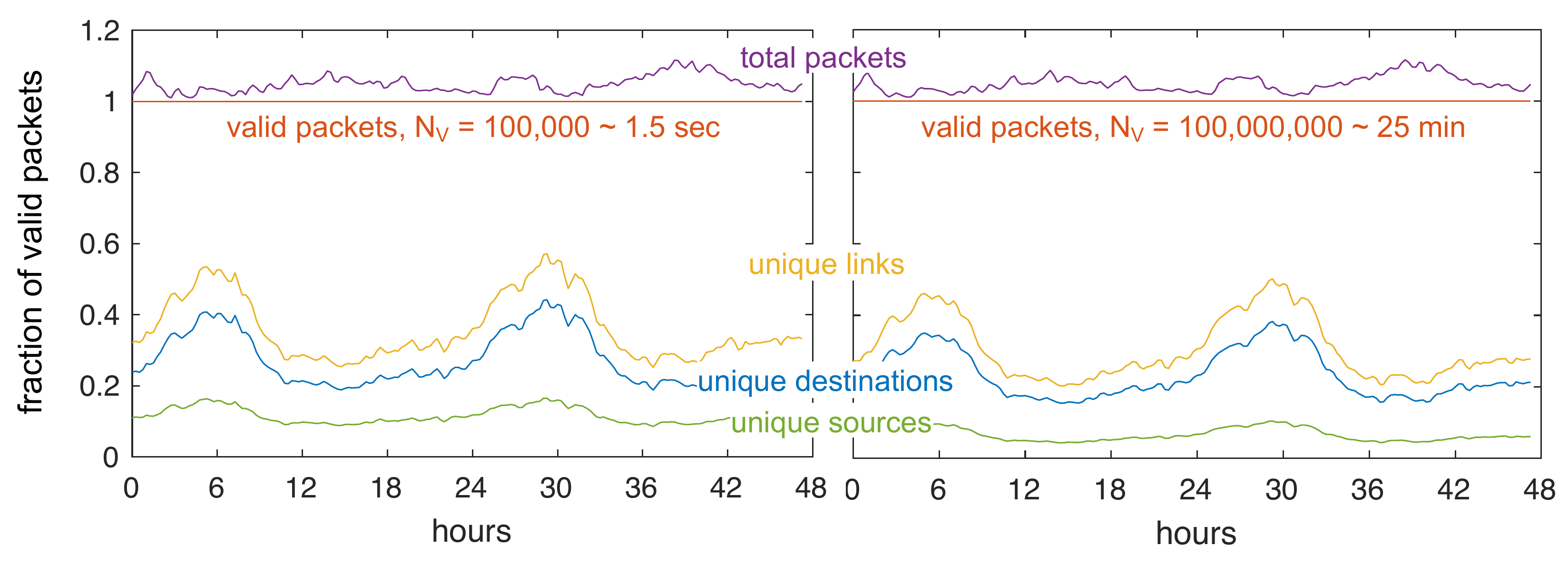}
	\includegraphics[width=\columnwidth]{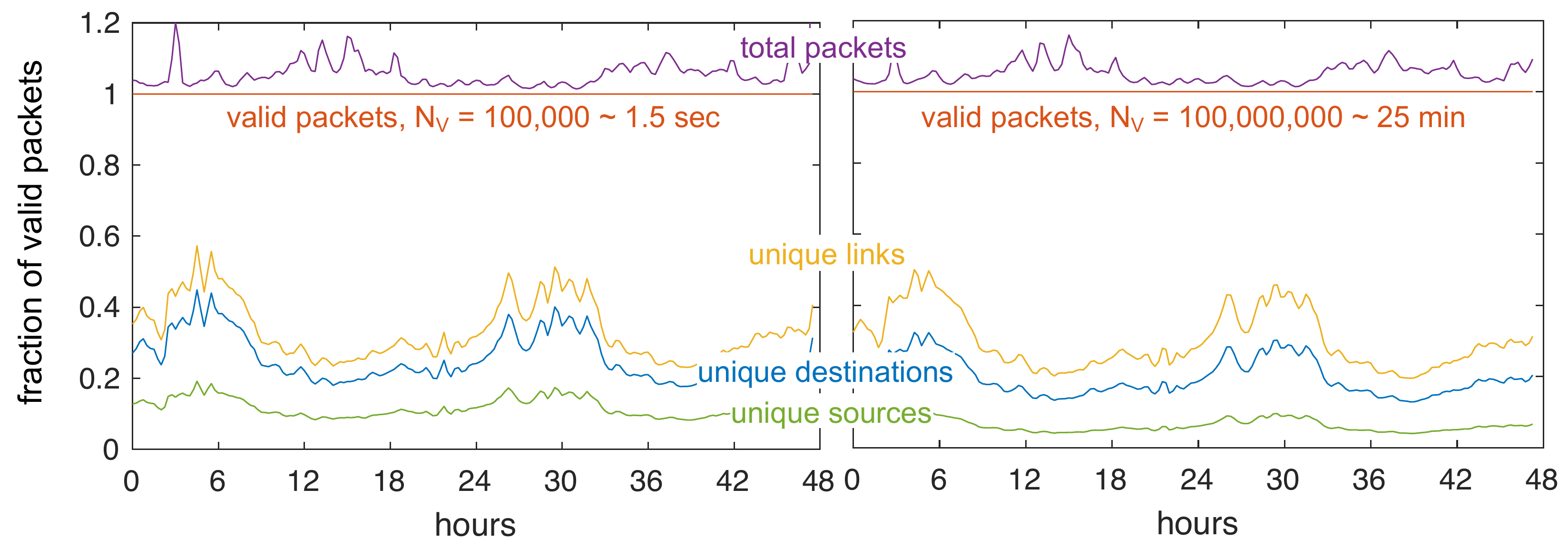}
	\caption{{\bf Valid packets.} Analyzing packet windows with the same numbers of valid packets produces consistent fractions of unique links, unique destinations, and unique sources over a wide range of packet sizes for the Tokyo 2015 (a) and Tokyo 2017 (b) data sets. The plots show these fractions for moving packet windows of  $N_V$ = 100{,}000 packets (left) and $N_V$ = 100{,}000{,}000 packets (right).  The packet windows correspond to time windows of approximately 1.5 seconds and 25 minutes.
	}
	\label{fig:ValidPackets}
\end{figure}

An essential step for increasing the accuracy of the statistical measures of Internet traffic is using windows with the same number of valid packets $N_V$.  For this analysis, a valid packet is defined as TCP over IPv4, which includes more than 95\% of the data in the collection and eliminates a small amount of data that use other protocols or contain anomalies.  Using packet windows with the same number of valid packets produces quantities that are consistent over a wide range from $N_V = 100{,}000$ to $N_V = 100{,}000{,}000$ (Figure~\ref{fig:ValidPackets}).

\subsection{Memory and Computation Requirements}

Processing 50 billion Internet packets with a variety of algorithms presents numerous computational challenges.  Dividing the data set into combinable units of approximately 100{,}000 consecutive packets made the analysis amenable to processing on a massively parallel supercomputer.  The detailed architecture of the parallel processing system and its corresponding performance are described in \cite{gadepally2018hyperscaling}.  The resulting processing pipeline was able to efficiently use over 10{,}000 processors on the MIT SuperCloud and was essential to this first-ever complete analysis of these data.

 A key element of our analysis is the use of novel sparse matrix mathematics in concert with the MIT SuperCloud.  Construction and analysis of network traffic matrices of the entire Internet address space have been considered impractical for its massive size \cite{tune2013internet}.  Internet Protocol version 4 (IPv4) has $2^{32}$ unique addresses, but at any given collection point, only a fraction of these addresses will be observed.  Exploiting this property to save memory can be accomplished by extending traditional sparse matrices so that new rows and columns can be added dynamically.

\begin{figure}
	\includegraphics[width=\columnwidth]{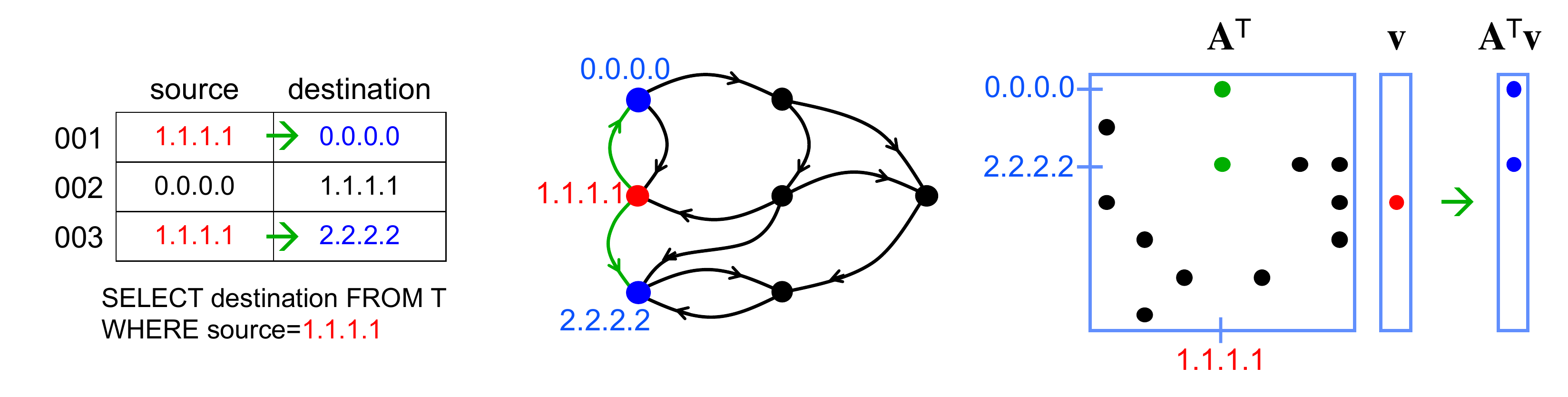}
	\caption{{\bf Associative arrays.} (a) Tabular representation of raw network traffic and corresponding database query to find all records beginning with source 1.1.1.1. (b) Network graph highlighting nearest neighbors of source node 1.1.1.1.  (c) Corresponding associative array representation of the network graph illustrating how the neighbors of source node 1.1.1.1 are computed with matrix vector multiplication.}
	\label{fig:AssociativeArrays}
\end{figure}

The algebra of associative arrays \cite{kepner2018mathematics} and its corresponding implementation in the Dynamic Distributed Dimensional Data Model (D4M) software library (d4m.mit.edu) allows the row and columns of a sparse matrix to be any sortable value, in this case character string representations of the Internet addresses (Figure~\ref{fig:AssociativeArrays}).  Associative arrays extend sparse matrices to have database table properties with dynamically insertable and removable rows and columns that adjust as new data are added or subtracted to the matrix.  Using these properties, the memory requirements of forming network traffic matrices can be reduced at the cost of increasing the required computation necessary to resort the rows and columns.

A network matrix ${\bf A}_t$ with $N_V = 100{,}000{,}000$ represented as an associative array typically requires 2 gigabytes of memory. A complete analysis of the statistics and topologies of ${\bf A}_t$ typically takes 10 minutes on a single MIT SuperCloud Intel Knights Landing processor core.   Using increments of $100{,}000$ packets means that this analysis is repeated over 500{,}000 times to process all 49.6 billion packets.  Using 10{,}000 processors on the MIT SuperCloud shortens the runtime of one of these analyses to approximately 8 hours.  The results presented within this chapter are products of a discovery process that required hundreds of such runs that would not have been possible without these computational resources.  Fortunately, the utilization of these results by Internet stakeholders can be significantly accelerated by creating optimized embedded implementations that only compute the desired statistics and are not required to support a discovery process \cite{liu2010tcam,liu2016packet}.

\section{Internet Traffic Modeling}
Quantitative measurements of the Internet \cite{rabinovich2016measuring} have provided Internet stakeholders  information on the Internet since its inception.  Early work has explored the early growth of the Internet \cite{claffy1994tracking}, the distribution of packet arrival times \cite{paxson1995wide}, the power-law distribution of network outages \cite{paxson1996end}, the self-similar behavior of traffic \cite{leland1994self,willinger1997self,willinger2002scaling}, formation processes of power-law networks \cite{faloutsos1999power,medina2000origin,broder2000graph,willinger2009mathematics}, and the topologies of Internet service providers \cite{spring2002measuring}.  Subsequent work has examined the technological properties of Internet topologies \cite{li2004first}, the diameter of the Internet \cite{leskovec2005graphs}, applying rank index-based Zipf--Mandelbrot modeling to peer-to-peer traffic \cite{saleh2006modeling}, and extending topology measurements to edge hosts \cite{heidemann2008census}.  More recent work looks to continued measurement of power-law phenomena \cite{mahanti2013tale,kitsak2015long,lischke2016analyzing}, exploiting emerging topologies for optimizing network traffic \cite{dhamdhere2010internet,labovitz2011internet,chiu2015we}, using network data to locate disruptions \cite{fontugne2017pinpointing}, the impact of inter-domain congestion \cite{dhamdhere2018inferring}, and studying the completeness of passive sources to determine how well they can observe microscopic phenomena \cite{mirkovic2017you}.

\begin{figure}
	\includegraphics[width=\columnwidth]{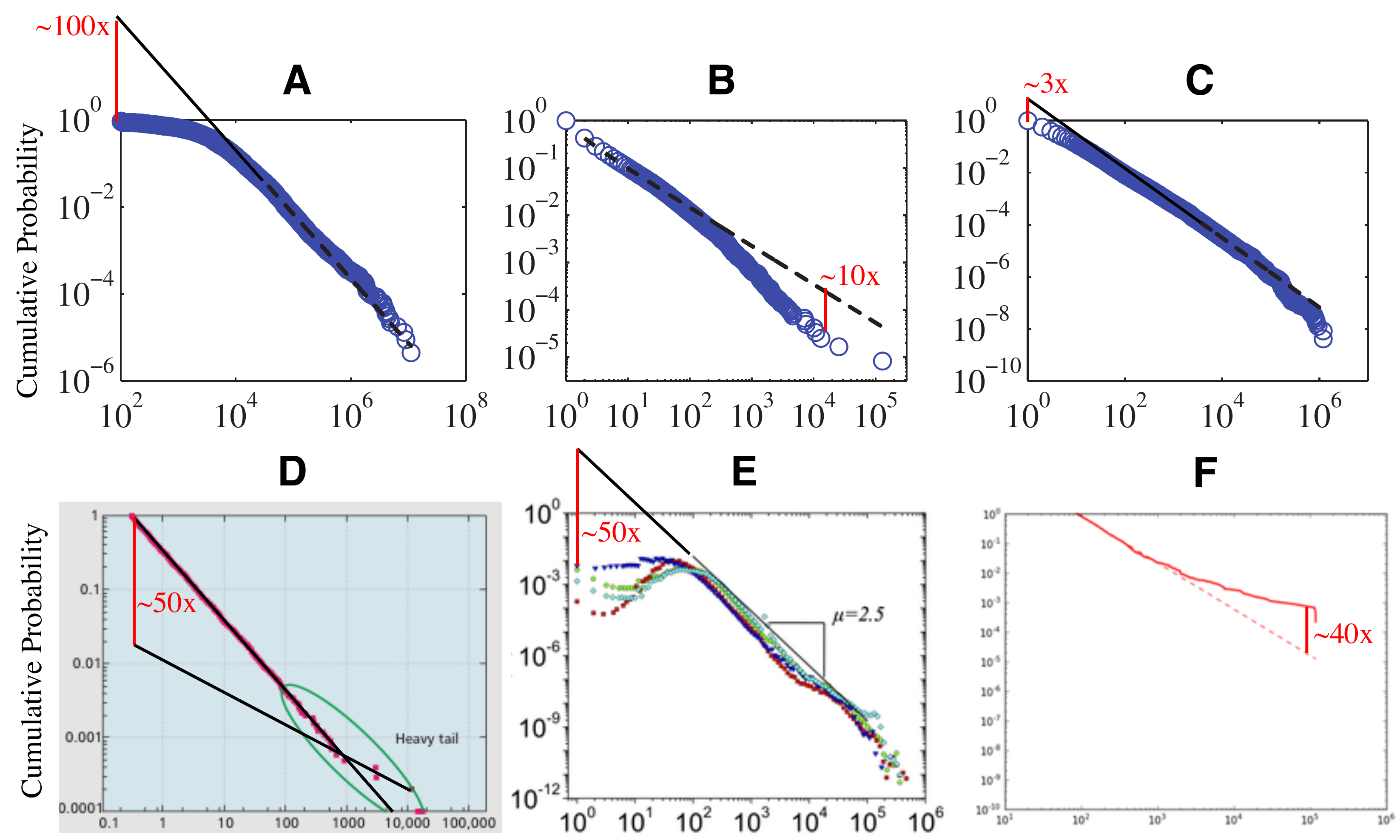}
	\caption{{\bf Single-parameter power-law fits of Internet data.} Single-parameter fits of the cumulative distributions of Internet data have difficulty modeling the entire range.  The estimated ratio between the model and the data at the model extremes is shown.
		({a}) Figure 8H \cite{clauset2009power}. ({b})  Figure 9W \cite{clauset2009power}.  ({c}) Figure 9X \cite{clauset2009power}.  ({d}) Figure 4B \cite{mahanti2013tale}.  ({e}) Figure 3A \cite{kitsak2015long}. ({f}) Figure 21 \cite{lischke2016analyzing}.}
	\label{fig:PowerLawFits}
\end{figure}

The above sample of many years of Internet research has provided significant qualitative insights into Internet phenomenology. Single-parameter power-law fits have extensively been explored and shown to adequately fit higher-degree tails of the observations.  However, more complex models are required to fit the entire range of observations. Figure~\ref{fig:PowerLawFits}a adapted from figure 8H \cite{clauset2009power} shows the number of bytes of data received in response to $2.3\times10^5$ HTTP (web) requests from computers at a large research laboratory and shows a strong agreement with a power law at large values, but diverges with the single-parameter model at small values. Figure~\ref{fig:PowerLawFits}b adapted from figure 9W \cite{clauset2009power} shows the distribution of $1.2\times10^5$ hits on web sites from AOL users and shows a strong agreement with a power law at small values, but diverges with the single-parameter model at large values.  Figure~\ref{fig:PowerLawFits}c adapted from figure 9X \cite{clauset2009power} shows the distribution of $2.4\times10^8$ web hyperlinks and has a reasonable model agreement across the entire range, except for the smallest values.  Figure~\ref{fig:PowerLawFits}d adapted from figure 4B \cite{mahanti2013tale} shows the distribution of visitors arriving at YouTube from referring web sites appears to be best represented by two very different power-law models with significant difference as the smallest values.  Figure~\ref{fig:PowerLawFits}e adapted from figure 3A \cite{kitsak2015long} shows the distribution of the number of Border Gateway Protocol updates received by the 4 monitors in 1-minute intervals and shows a strong agreement with a power law at large values, but diverges with the single-parameter model at small values. Figure~\ref{fig:PowerLawFits}f adapted from figure 21 in \cite{lischke2016analyzing} shows the distribution of the Bitcoin network in 2011 and shows a strong agreement with a power law at small values, but diverges with the single-parameter model at large values.

The results shown in Figure~\ref{fig:PowerLawFits} represent some of the best and most carefully executed fits to Internet data and clearly show the difficulty of fitting the entire range with a single-parameter power law. It is also worth mentioning that in each case the cumulative distribution is used, which naturally provides a smoother curve (in contrast to the differential cumulative distribution used in our analysis), but provides less detail on the underlying phenomena. Furthermore, the data in Figure~\ref{fig:PowerLawFits} are typically isolated collections such that the error bars are not readily computable, which limits the ability to assess both the quality of the measurements and the model fits.

Regrettably, the best publicly available data about the global interconnection system that carries most of the world's communications traffic are incomplete and of unknown accuracy.  There is no map of physical link locations, capacity, traffic, or interconnection arrangements.  This opacity of the Internet infrastructure hinders research and development efforts to model network behavior and topology; design protocols and new architectures; and study real-world properties such as robustness, resilience, and economic sustainability.  There are good reasons for the dearth of information: complexity and scale of the infrastructure; information-hiding properties of the routing system; security and commercial sensitivities; costs of storing and processing the data; and lack of incentives to gather or share data in the first place, including cost-effective ways to use it operationally. But understanding the Internet's history and present, much less its future, is impossible without realistic and representative data sets and measurement infrastructure on which to support sustained longitudinal measurements as well as new experiments.  The MAWI and CAIDA data collection efforts are the  largest efforts to provide the data necessary to begin to answer these questions.

\subsection{Logarithmic Pooling}\label{prob}
\begin{figure}
	\centering
	\includegraphics[width=0.75\columnwidth]{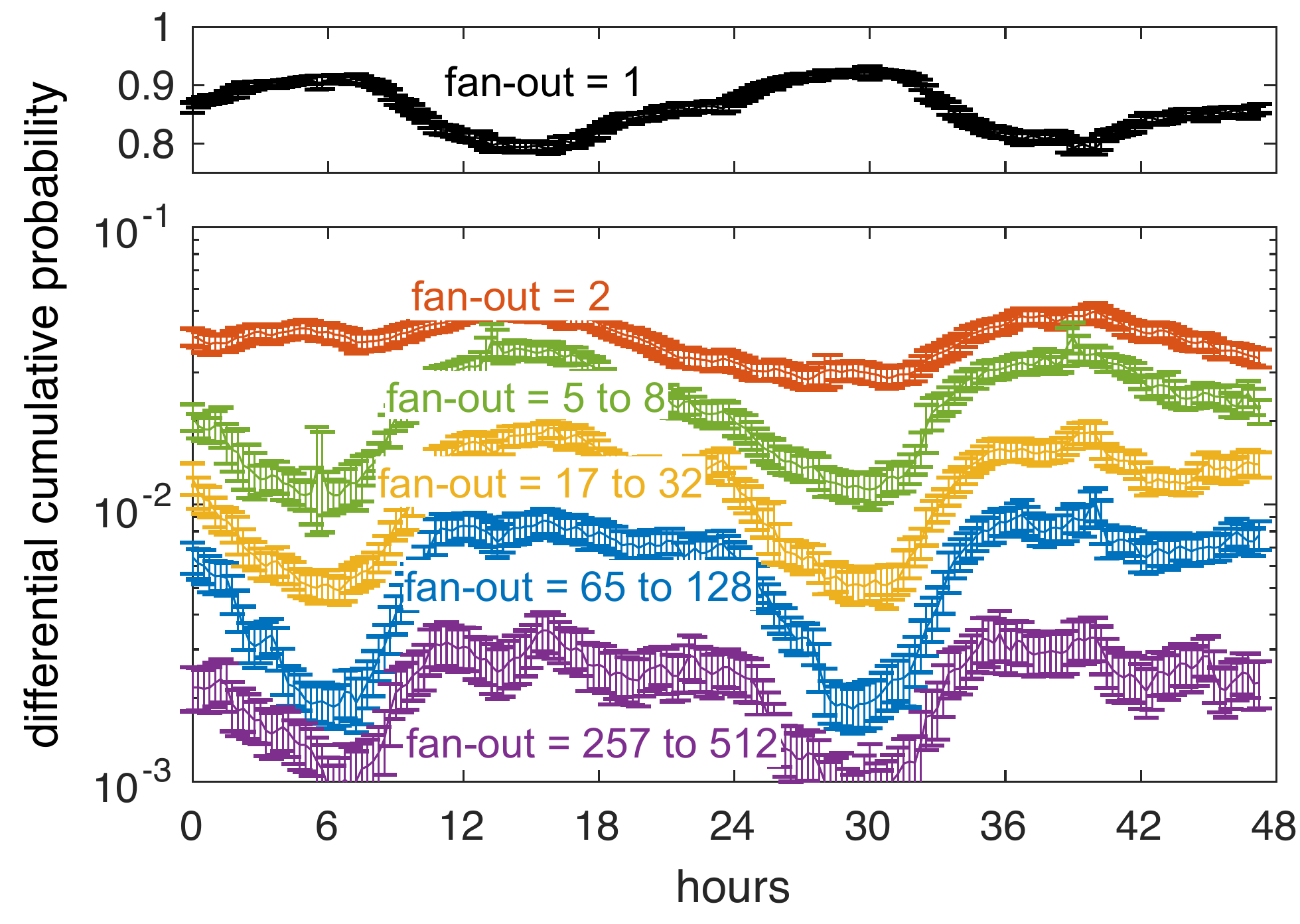}
	\caption{{\bf Daily variation in Internet traffic.}  The fraction of source nodes with a given range of fan-out is shown as a function of time for the Tokyo 2015 data.  The $p(d = 1)$ value is plotted on a separate linear scale because of the larger magnitude relative to the other points.  Each point is the mean of many neighboring points in time, and the error bars are the measured $\pm$1-$\sigma$.  The daily variations of the distributions oscillate between extremes corresponding to approximately local noon and midnight.
	}
	\label{fig:DailyVariation}
\end{figure}

\begin{figure}
	\centering
	\includegraphics[width=0.75\columnwidth]{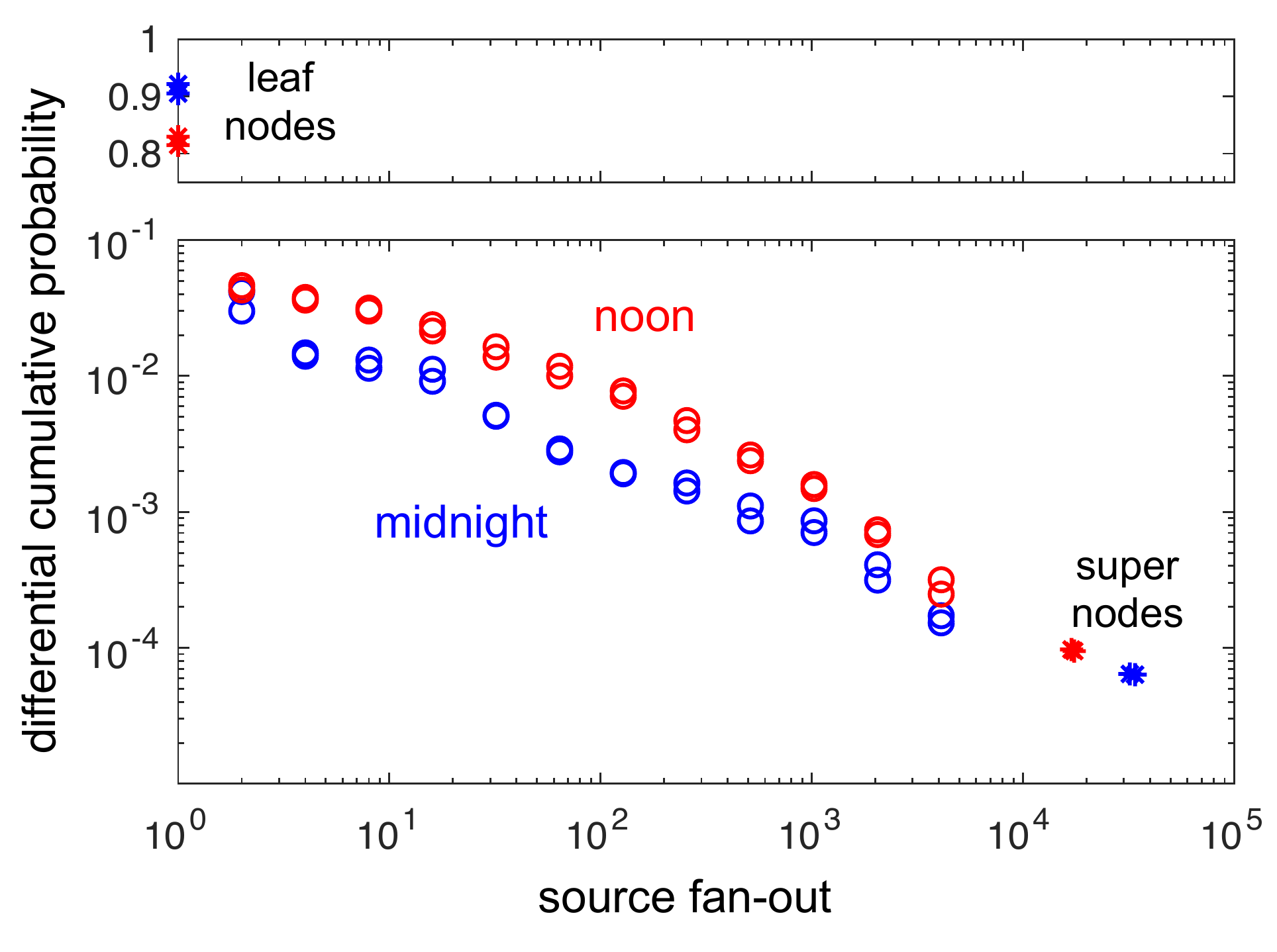}
	\caption{{\bf Daily limits in Internet traffic.}  The fraction of source nodes versus fan-out is shown for two noons and two midnights for the Tokyo 2015 data.  The overlap among the noons and the midnights shows the relative day-to-day consistency in these data and shows the limits of the two extremes in daily variation.  During the day, there is more traffic among nodes with intermediate fan-out.  At night, the traffic is more dominated by leaf nodes and the supernode.
	}
	\label{fig:DailyLimits}
\end{figure}

In this analysis before model fitting, the differential cumulative probabilities are calculated.
For a network quantity $d$, the histogram of this quantity computed from ${\bf A}_t$ is denoted by $n_t(d)$, with corresponding probability
\begin{equation}\label{eq:Probability}
p_t(d) = n_t(d)/\sum_d n_t(d)
\end{equation}
and cumulative probability
\begin{equation}\label{eq:Cumulative}
P_t(d) = \sum_{i=1,d} p_t(d)
\end{equation}
Because of the relatively large values of $d$ observed due to a single supernode, the measured probability at large $d$ often exhibits large fluctuations. However, the cumulative probability lacks sufficient detail to see variations around specific values of $d$, so it is typical to use the differential cumulative probability with logarithmic bins in $d$
\begin{equation}\label{eq:LogBin}
D_t(d_i) = P_t(d_i) - P_t(d_{i-1})
\end{equation}
where $d_i = 2^i$ \cite{clauset2009power}.  The corresponding mean and standard deviation of $D_t(d_i)$ over many different consecutive values of $t$ for a given data set are denoted $D(d_i)$ and $\sigma(d_i)$. These quantities strike a balance between accuracy and detail for subsequent model fitting as demonstrated in the daily structural variations revealed in the Tokyo data (Figures~\ref{fig:DailyVariation} and \ref{fig:DailyLimits}).


Diurnal variations in supernode network traffic are well known \cite{soule2004identify}. The Tokyo packet data were collected over a period spanning two days and allow the daily variations in packet traffic to be observed.  The precision and accuracy of our measurements allow these variations to be observed across a wide range of nodes. Figure~\ref{fig:DailyVariation} shows the fraction of source fan-outs in each of various bin ranges.  The fluctuations show the network evolving between two envelopes occurring  between noon and midnight that are shown in Figure~\ref{fig:DailyLimits}.

\subsection{Modified Zipf--Mandelbrot Model}\label{zipf}
Measurements of $D(d_i)$ can reveal many properties of network traffic, such as the number of nodes with only one connection $D(d = 1)$ and the size of the supernode $d_{\rm max}={\rm argmax}(D(d) > 0)$. An effective low-parameter model allows these and many other properties to be summarized and computed efficiently.  In the standard Zipf--Mandelbrot model typically used in linguistic contexts, the value $d$ in Eq.~\ref{eq:ZipfMandelbrot} is a ranking with $d=1$ corresponding to the most popular value \cite{mandelbrot1953informational,montemurro2001beyond,saleh2006modeling}.  In our analysis, the Zipf--Mandelbrot model is modified so that $d$ is a measured network quantity instead of a rank index (Eq. \ref{eq:ZipfMandelbrot}).  The model exponent $\alpha$ has a larger impact on the model at large values of $d$, while the model offset $\delta$ has a larger impact on the model at small values of $d$ and in particular at $d=1$.

The general saturation/cutoff models used to model a variety of network phenomena is denoted \cite{clauset2009power,barabasi2016network}
\begin{equation}\label{eq:satcut}
p(d) \propto \frac{1}{(d + \delta)^\alpha \exp[\lambda d]}
\end{equation}
where $\delta$ is the low-$d$ saturation and $1/\lambda$ is the high-$d$ cutoff that bounds the power-law regime of the distribution.  The modified Zipf--Mandelbrot is a special case of this distribution that accurately models our observations. The unnormalized modified Zipf--Mandelbrot model is denoted
\begin{equation}\label{eq:rho}
\rho(d;\alpha,\delta) = \frac{1}{(d + \delta)^\alpha}
\end{equation}
with corresponding derivative with respect to $\delta$
\begin{equation}\label{eq:drho}
\partial_\delta \rho(d;\alpha,\delta) = \frac{-\alpha}{(d + \delta)^{\alpha+1}} = -\alpha \rho(d;\alpha+1,\delta)
\end{equation}
The normalized model probability is given by
\begin{equation}\label{eq:ZM}
p(d;\alpha,\delta) = \frac{\rho(d;\alpha,\delta)}{\sum_{d=1}^{d_{\rm max}} \rho(d;\alpha,\delta)}
\end{equation}
where $d_{\rm{max}}$ is the largest value of the network quantity $d$.  The cumulative model probability is the sum
\begin{equation}\label{eq:ZMcum}
P(d_i;\alpha,\delta) = \sum_{d=1}^{d_i} p(d;\alpha,\delta)
\end{equation}
The corresponding differential cumulative model probability is
\begin{equation}\label{eq:ZMdiff}
D(d_i;\alpha,\delta) = P(d_i;\alpha,\delta) - P(d_{i-1};\alpha,\delta)
\end{equation}
where $d_i = 2^i$.  In terms of $\rho$, the differential cumulative model probability is
\begin{equation}\label{eq:ZMdiffrho}
D(d_i;\alpha,\delta) = \frac{\sum_{d=d_{i-1}+1}^{d=d_i} \rho(d;\alpha,\delta)}{\sum_{d=1}^{d = d_{\rm max}}~~\rho(d;\alpha,\delta)}
\end{equation}
The above function is closely related to the Hurwitz zeta function \cite{NIST:DLMF,clauset2009power,yu2017link}  \begin{equation}\label{eq:HZ}
\zeta(\alpha,\delta_1) = \sum_{d=0}^{\infty}~~\rho(d;\alpha,\delta_1)
\end{equation}
where $\delta_1 = \delta+1$.  The differential cumulative model probability in terms of the Hurwitz zeta function is
\begin{equation}\label{eq:ZMdiffrho}
D(d_i;\alpha,\delta) = \frac{\zeta(\alpha,\delta+3+d_{i-1}) - \zeta(\alpha,\delta+2+d_i)}{\zeta(\alpha,\delta+) - \zeta(\alpha,\delta+)}
\end{equation}

\subsection{Nonlinear Model Fitting}\label{nonlinear}
The model exponent $\alpha$ has a larger impact on the model at large values of $d$, while the model offset $\delta$ has a larger impact on the model at small values of $d$ and in particular at $d=1$. A nonlinear fitting technique is used to obtain accurate model fits across the entire range of $d$.  Initially, a set of candidate exponent values is selected, typically $\alpha = 0.10, 0.11,\ldots,3.99,4.00$. For each value of $\alpha$, a value of $\delta$ is computed that exactly matches the model with the data at $D(1)$.  Finding the value of $\delta$ corresponding to a give $D(1)$ is done using Newton's method as follows.  Setting the measured value of $D(1)$  equal to the model value $D(1;\alpha,\delta)$ gives
\begin{equation}\label{eq:ZM1}
D(1) = D(1;\alpha,\delta) = \frac{1}{(1 + \delta)^{\alpha} \sum_{d=1}^{d_{\rm max}} \rho(d;\alpha,\delta)}
\end{equation}
Newton's method works on functions of the form $f(\delta) = 0$. Rewriting the above expression produces
\begin{equation}\label{eq:ZMnewton}
f(\delta) = D(1) (1 + \delta)^\alpha \sum_{d=1}^{d_{\rm max}} \rho(d;\alpha,\delta) - 1 = 0
\end{equation}
For given value of $\alpha$, $\delta$ can be computed from the following iterative equation
\begin{equation}\label{eq:NewtonIteration}
\delta \rightarrow \delta - \frac{f(\delta)}{\partial_\delta f(\delta)}
\end{equation}
where the partial derivative $\partial_\delta f(\delta)$ is
\begin{eqnarray}\label{eq:NewtonDerivative}
\partial_\delta f(\delta)
& = & D(1) ~ \partial_\delta [(1 + \delta)^\alpha ~~ \sum_{d=1}^{d_{\rm max}} \rho(d;\alpha,\delta)] \nonumber \\
& = & D(1) [[\alpha (1 + \delta)^{\alpha-1} \sum_{d=1}^{d_{\rm max}} \rho(d;\alpha,\delta)] +
[(1 + \delta)^\alpha \sum_{d=1}^{d_{\rm max}} ~~ \partial_\delta \rho(d;\alpha,\delta)]]
\nonumber \\
& = & D(1) [[\alpha (1 + \delta)^{\alpha-1} \sum_{d=1}^{d_{\rm max}} \rho(d;\alpha,\delta)] +
[(1 + \delta)^\alpha \sum_{d=1}^{d_{\rm max}} -\alpha \rho(d;\alpha+1,\delta)]] \nonumber \\
& = & \alpha D(1) (1 + \delta)^\alpha [(1 + \delta)^{-1} \sum_{d=1}^{d_{\rm max}} \rho(d;\alpha,\delta) -
\sum_{d=1}^{d_{\rm max}} \rho(d;\alpha+1,\delta)]
\end{eqnarray}
Using a starting value of $\delta=1$ and bounds of $0 < \delta < 10$, Newton's method can be iterated until the differences in successive values of $\delta$ fall below a specified error (typically 0.001), which is usually achieved in less than five iterations.

If faster evaluation is required, the sums in the above formulas can be accelerated using the integral approximations
\begin{eqnarray}\label{eq:SumApprox}
\sum_{d=1}^{d_{\rm max}} \rho(d;\alpha,\delta)
&\approx& \sum_{d=1}^{d_{\rm sum}} \rho(d;\alpha,\delta) + \int_{d_{\rm sum} + 0.5}^{d_{\rm max} + 0.5} \rho(x;\alpha,\delta) dx \nonumber \\
&=& \sum_{d=1}^{d_{\rm sum}} \rho(d;\alpha,\delta) + \frac{\rho(d_{\rm sum} + 0.5;\alpha-1,\delta) - \rho(d_{\rm max} + 0.5;\alpha-1,\delta) }{\alpha-1} \nonumber \\
\sum_{d=1}^{d_{\rm max}} \rho(d;\alpha+1,\delta)
&\approx& \sum_{d=1}^{d_{\rm sum}} \rho(d;\alpha+1,\delta) + \int_{d_{\rm sum} + 0.5}^{d_{\rm max} + 0.5} \rho(x;\alpha+1,\delta) dx \nonumber \\
&=& \sum_{d=1}^{d_{\rm sum}} \rho(d;\alpha+1,\delta) + \frac{\rho(d_{\rm sum} + 0.5;\alpha,\delta) - \rho(d_{\rm max} + 0.5;\alpha,\delta) }{\alpha}
\end{eqnarray}
where the parameter $d_{\rm sum}$ can be adjusted to exchange speed for accuracy.  For typical values of $\alpha$, $\delta$, and $d_{\rm max}$ used in this work, the accuracy is approximately $1/d_{\rm sum}$.

The best-fit $\alpha$ (and corresponding $\delta$) is chosen by minimizing the $|~|^{1/2}$ metric over logarithmic differences between the candidate models $D(d_i;\alpha,\delta)$ and the data
\begin{equation}\label{eq:NonLinFit}
{\rm argmin}_{\alpha} \sum_{d_i}|\log(D(d_i)) - \log(D(d_i;\alpha,\delta))|^{1/2}
\end{equation}
The $|~|^{1/2}$ metric (or $|~|_p$-norm with $p = 1/2$) favors maximizing error sparsity over minimizing outliers \cite{donoho2006compressed,chartrand2007exact,xu2012,karvanen2003measuring,saito2000sparsity,Brbic2018,Rahimi2018scale}. Several authors have recently shown that it is possible to reconstruct a nearly sparse signal from fewer linear measurements than would be expected from traditional sampling theory.  Furthermore, by replacing the $|~|_1$ norm with the $|~|^p$ with $p < 1$,  reconstruction is possible with substantially fewer measurements.

Using logarithmic values more evenly weights their contribution to the model fit and more accurately reflects the number of packets used to compute each value of $D(d_i)$.  Lower-accuracy data points are avoided by limiting the fitting procedure to data points where the value is greater than the standard deviation: $D(d_i) > \sigma(d_i)$.

\section{Results}\label{Results}
\begin{figure}
	\vspace*{-0.5cm}
	\hspace*{-1cm}
	\includegraphics[height = 19cm, width=1.1\columnwidth]{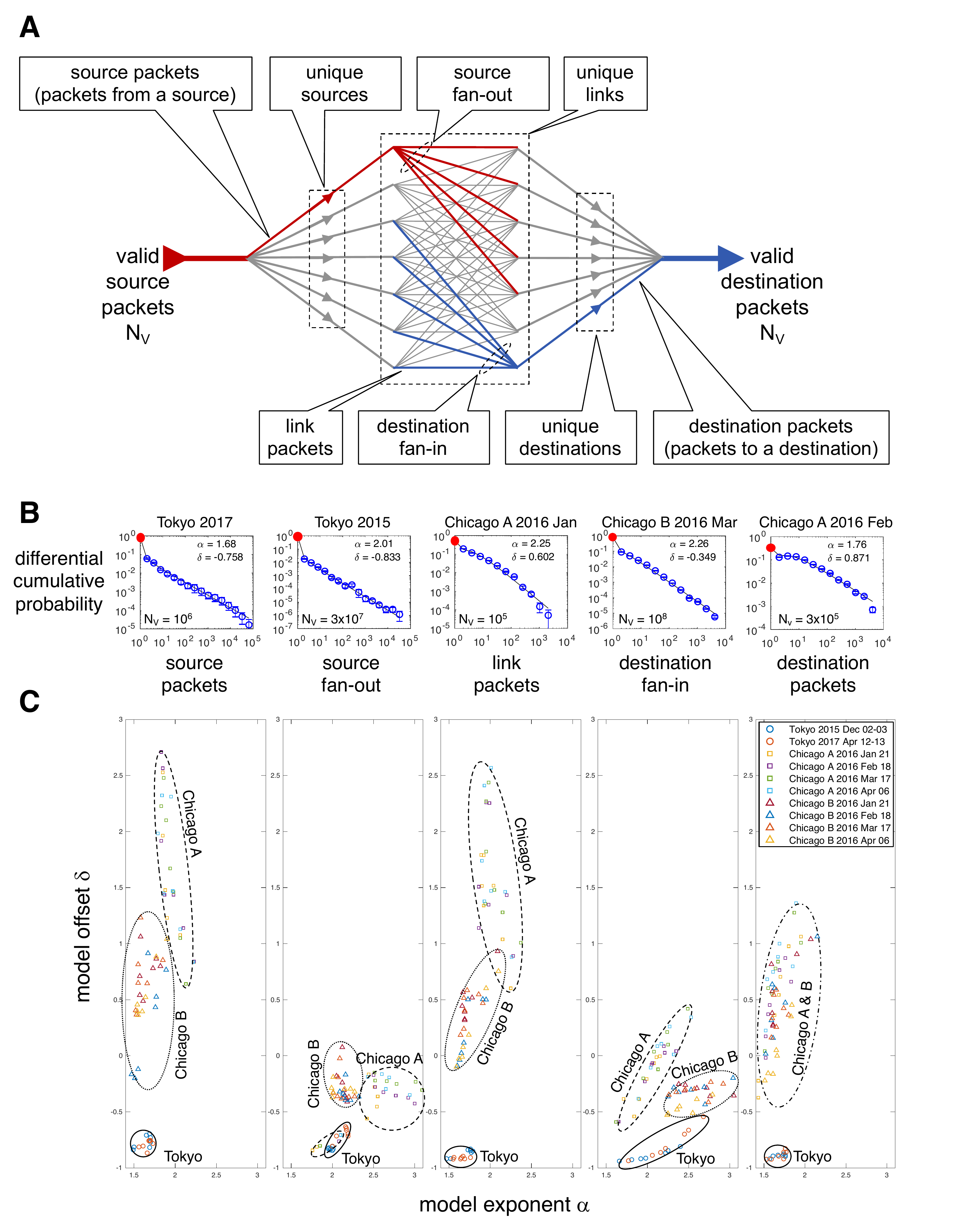}
	\vspace*{-1cm}
	\caption{{\bf Streaming network traffic quantities, distributions, and model fits.} ({a}) Internet traffic streams of $N_V$ valid packets are divided into a variety of quantities for analysis. ({b}) A selection of 5 of the 350  measured differential cumulative probabilities spanning different locations, dates, and packet windows.  Blue circles are measured data with $\pm$1-$\sigma$ error bars.  Black lines are the best-fit modified Zipf--Mandelbrot models with parameters $\alpha$ and $\delta$.  Red dots highlight the large contribution of leaf nodes and isolated links.  ({c}) Model fit parameters for all 350 measured probability distributions.}
	\label{fig:NetworkDistribution}
\end{figure}

\begin{figure}[bh]
	\vspace*{-0.5cm}
	\hspace*{-1cm}
	\vspace*{-.9cm}
    \includegraphics[clip, trim=2cm 0cm 0cm 0cm, angle=-90,origin=c,width=\columnwidth]{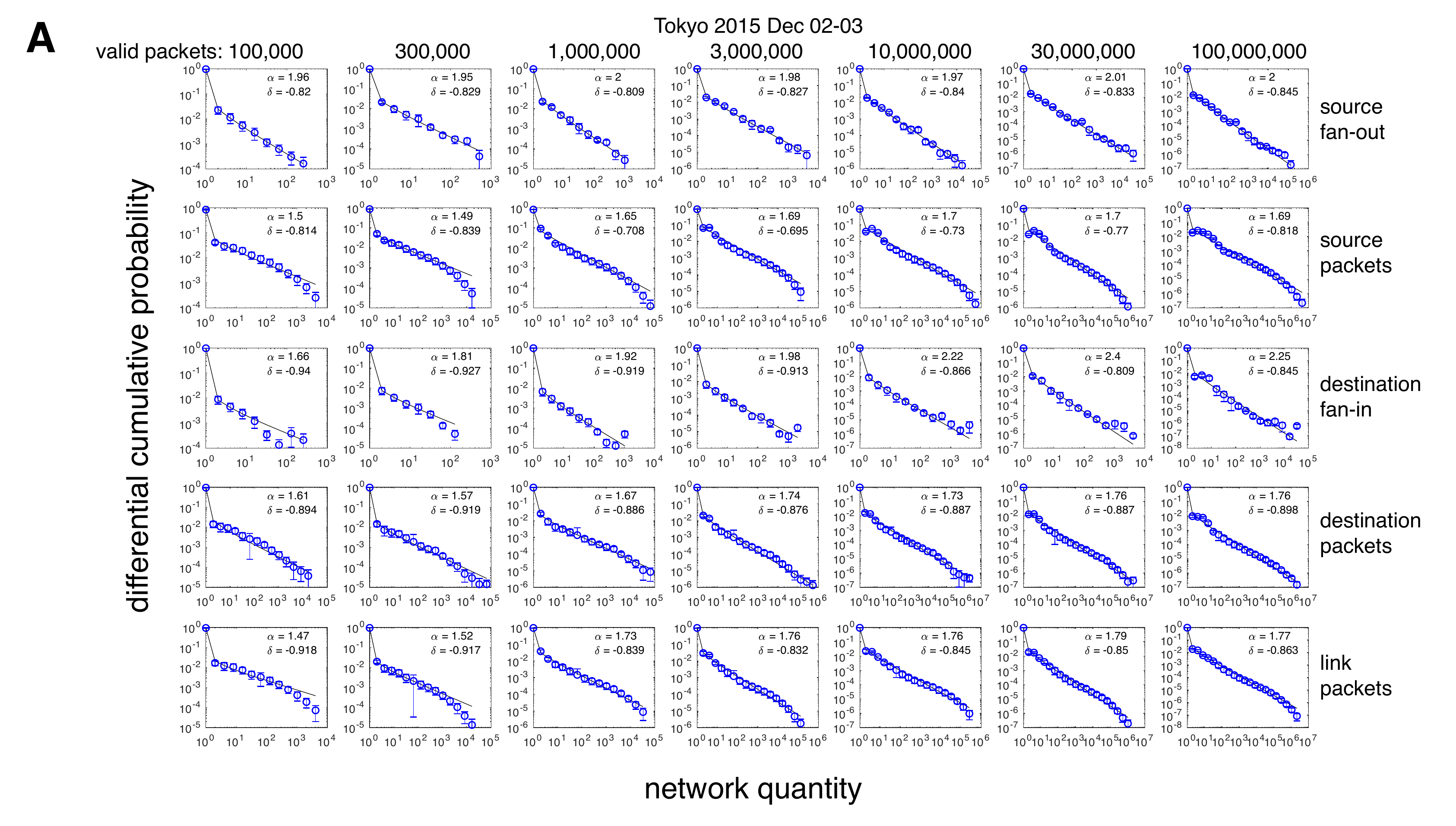}
	\caption{{\bf Measured differential cumulative probabilities.}  Blue circles with $\pm$1-$\sigma$ error bars, along with their best-fit modified Zipf--Mandelbrot models (black line) and parameters $\alpha$ and $\delta$ performed for Tokyo 2015 Dec 02.	
	}
	\label{fig:ModelFitsA}
\end{figure}

\begin{figure}[bh]
	\vspace*{-0.5cm}
	\hspace*{-1cm}
	\vspace*{-.9cm}
	\includegraphics[clip, trim=2cm 0cm 0cm 0cm, angle=-90,origin=c,width=\columnwidth]{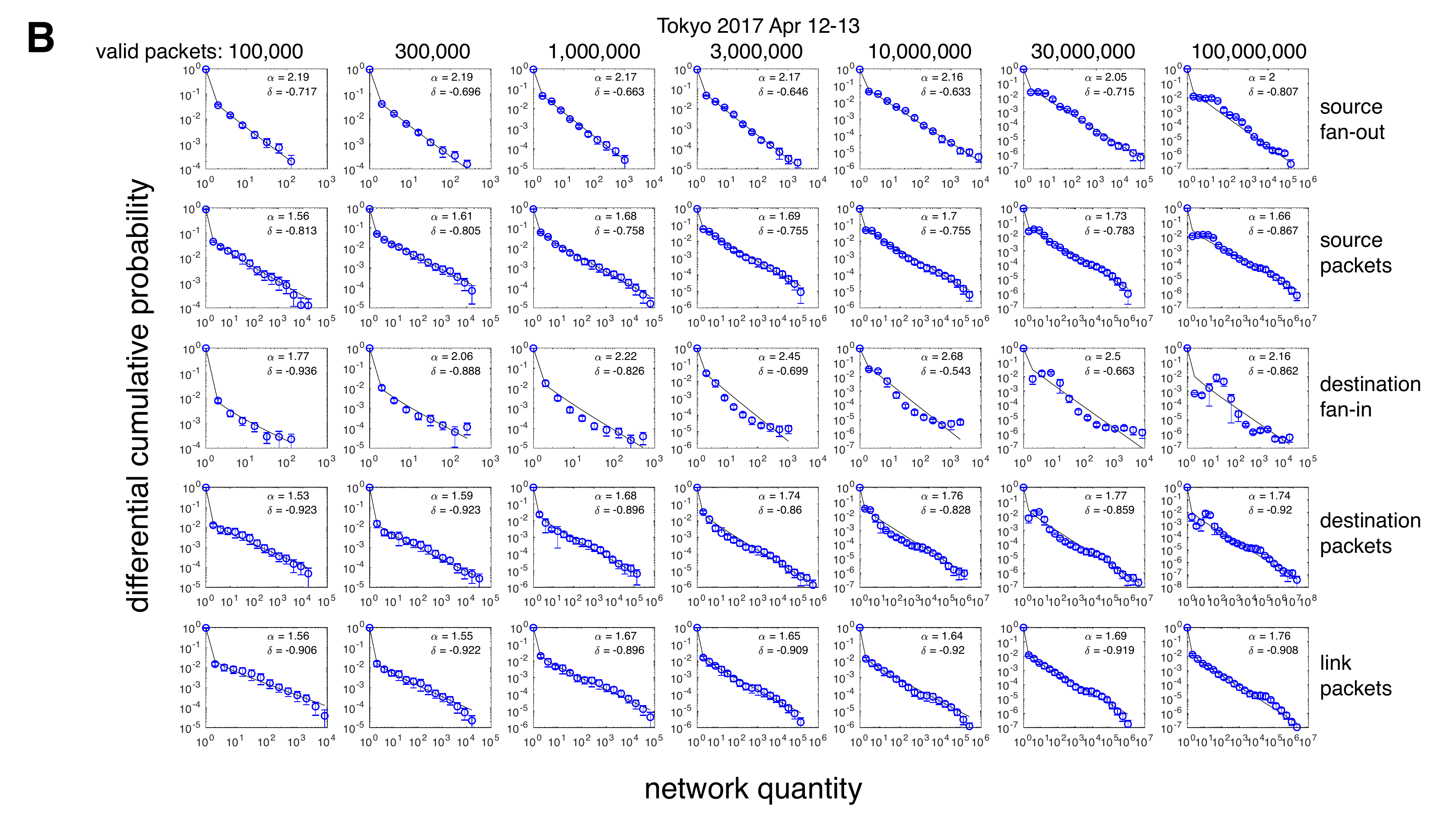}
	\caption{{\bf Measured differential cumulative probabilities.}  Blue circles with $\pm$1-$\sigma$ error bars, along with their best-fit modified Zipf--Mandelbrot models (black line) and parameters $\alpha$ and $\delta$ performed for Tokyo 2017 Apr 12.	
	}
	\label{fig:ModelFitsB}
\end{figure}

\begin{figure}[bh]
	\vspace*{-0.5cm}
	\hspace*{-1cm}
	\vspace*{-.9cm}
	\includegraphics[clip, trim=2cm 0cm 0cm 0cm, angle=-90,origin=c,width=\columnwidth]{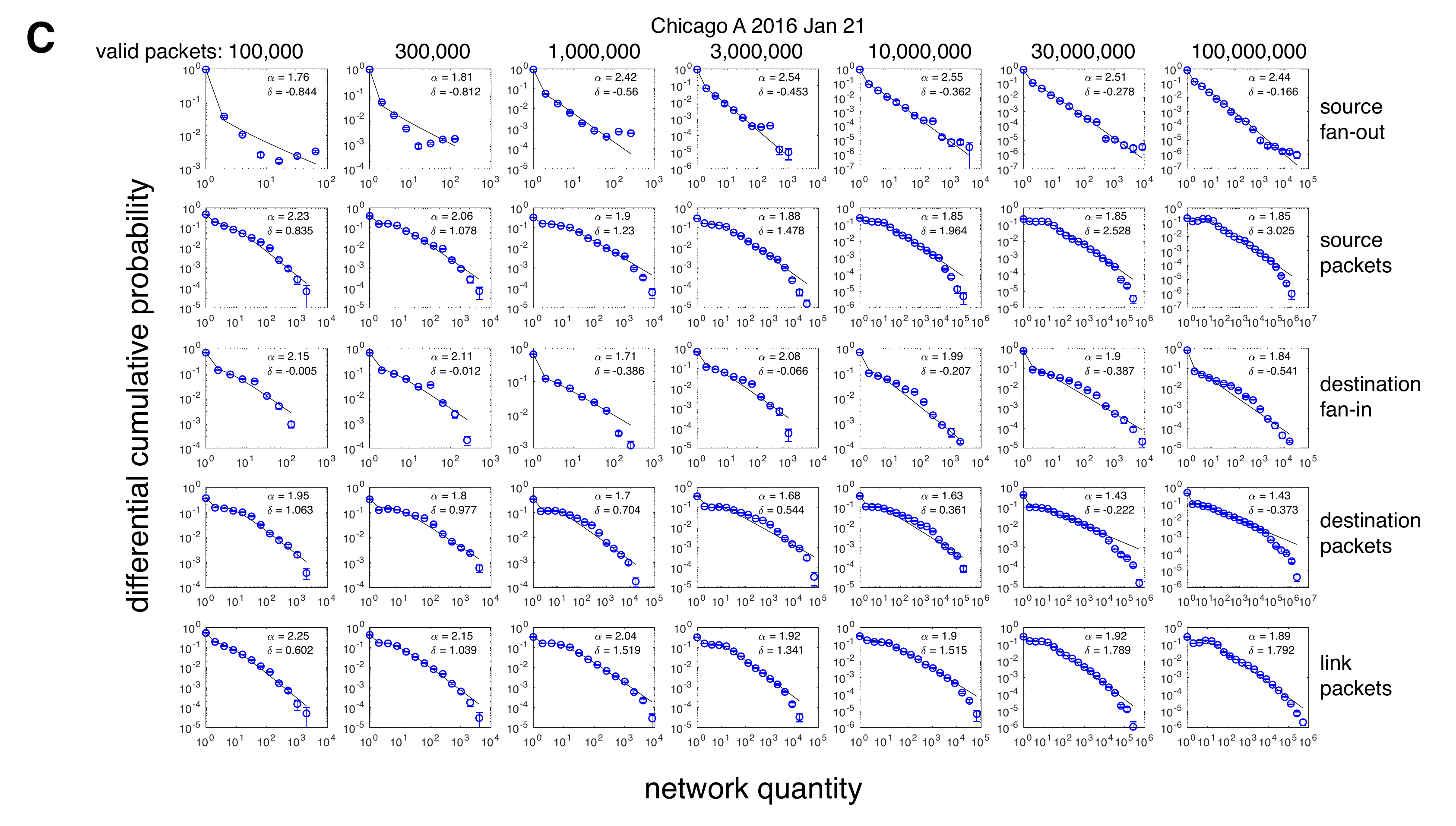}
	\caption{{\bf Measured differential cumulative probabilities.}  Blue circles with $\pm$1-$\sigma$ error bars, along with their best-fit modified Zipf--Mandelbrot models (black line) and parameters $\alpha$ and $\delta$ performed for Chicago A 2016 Jan 21.	
	}
	\label{fig:ModelFitsC}
\end{figure}

\begin{figure}[bh]
	\vspace*{-0.5cm}
	\hspace*{-1cm}
	\vspace*{-.9cm}
	\includegraphics[clip, trim=2cm 0cm 0cm 0cm, angle=-90,origin=c,width=\columnwidth]{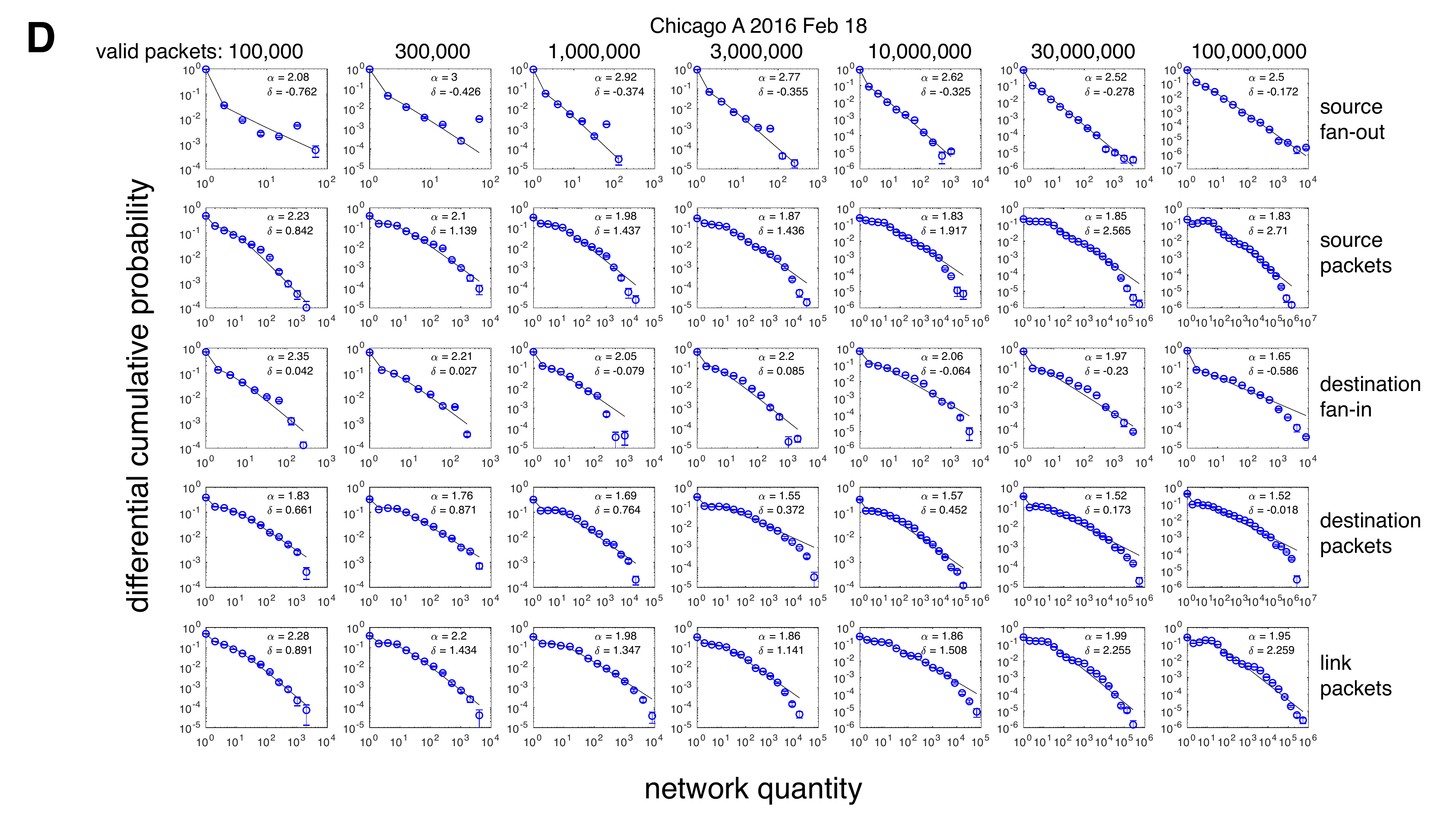}
	\caption{{\bf Measured differential cumulative probabilities.}  Blue circles with $\pm$1-$\sigma$ error bars, along with their best-fit modified Zipf--Mandelbrot models (black line) and parameters $\alpha$ and $\delta$ performed for Chicago A 2016 Feb 18.	
	}
	\label{fig:ModelFitsD}
\end{figure}

\begin{figure}[bh]
	\vspace*{-0.5cm}
	\hspace*{-1cm}
	\vspace*{-.9cm}
	\includegraphics[clip, trim=2cm 0cm 0cm 0cm, angle=-90,origin=c,width=\columnwidth]{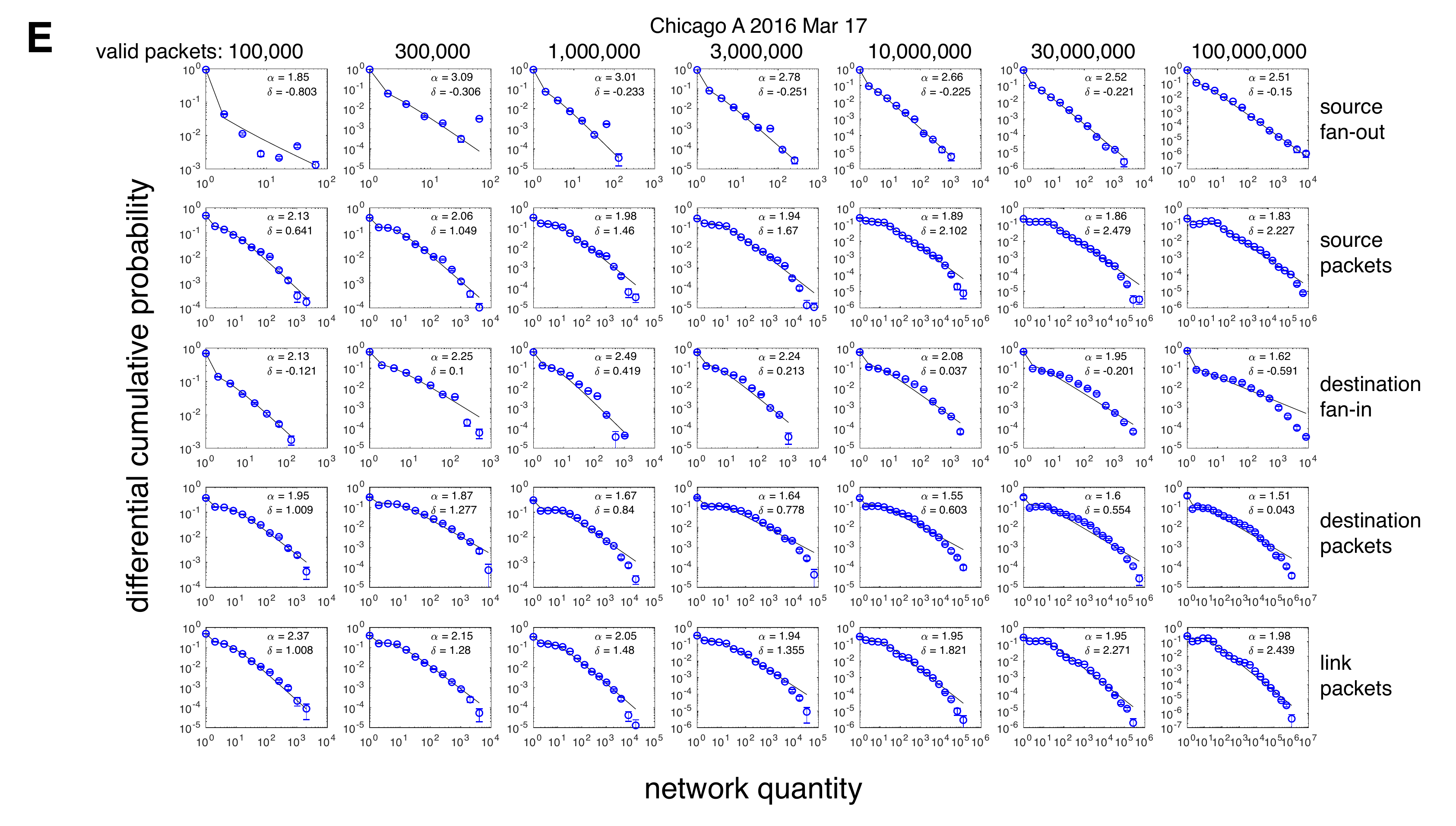}
	\caption{{\bf Measured differential cumulative probabilities.} Blue circles with $\pm$1-$\sigma$ error bars, along with their best-fit modified Zipf--Mandelbrot models (black line) and parameters $\alpha$ and $\delta$ performed for Chicago A 2016 Mar 17.	
	}
	\label{fig:ModelFitsE}
\end{figure}

\begin{figure}[bh]
	\vspace*{-0.5cm}
	\hspace*{-1cm}
	\vspace*{-.9cm}
	\includegraphics[clip, trim=2cm 0cm 0cm 0cm, angle=-90,origin=c,width=\columnwidth]{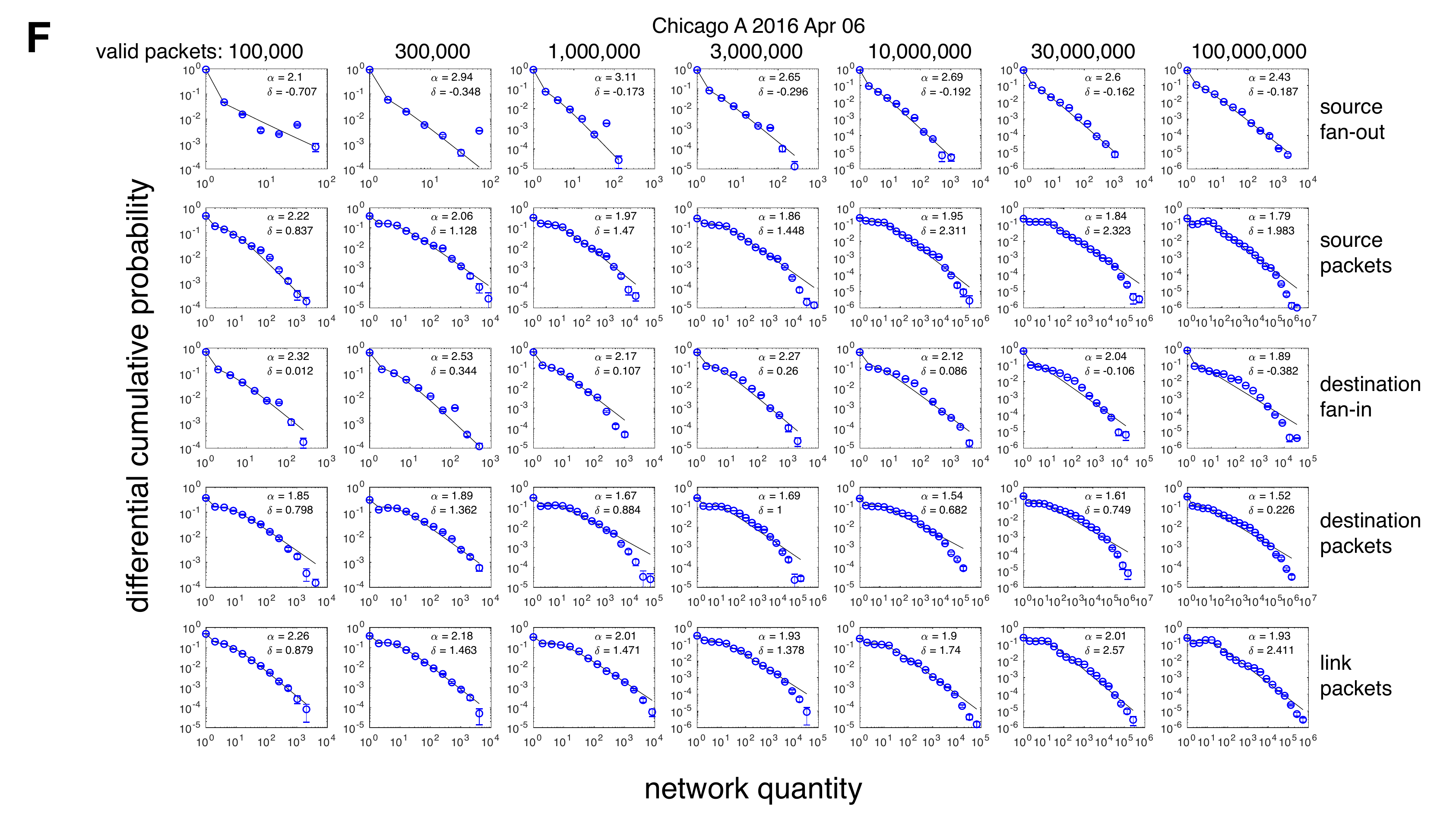}
	\caption{{\bf Measured differential cumulative probabilities.} Blue circles with $\pm$1-$\sigma$ error bars, along with their best-fit modified Zipf--Mandelbrot models (black line) and parameters $\alpha$ and $\delta$ performed for Chicago A 2016 Apr 06.	
	}
	\label{fig:ModelFitsF}
\end{figure}

\begin{figure}[bh]
	\vspace*{-0.5cm}
	\hspace*{-1cm}
	\vspace*{-.9cm}
	\includegraphics[clip, trim=2cm 0cm 0cm 0cm, angle=-90,origin=c,width=\columnwidth]{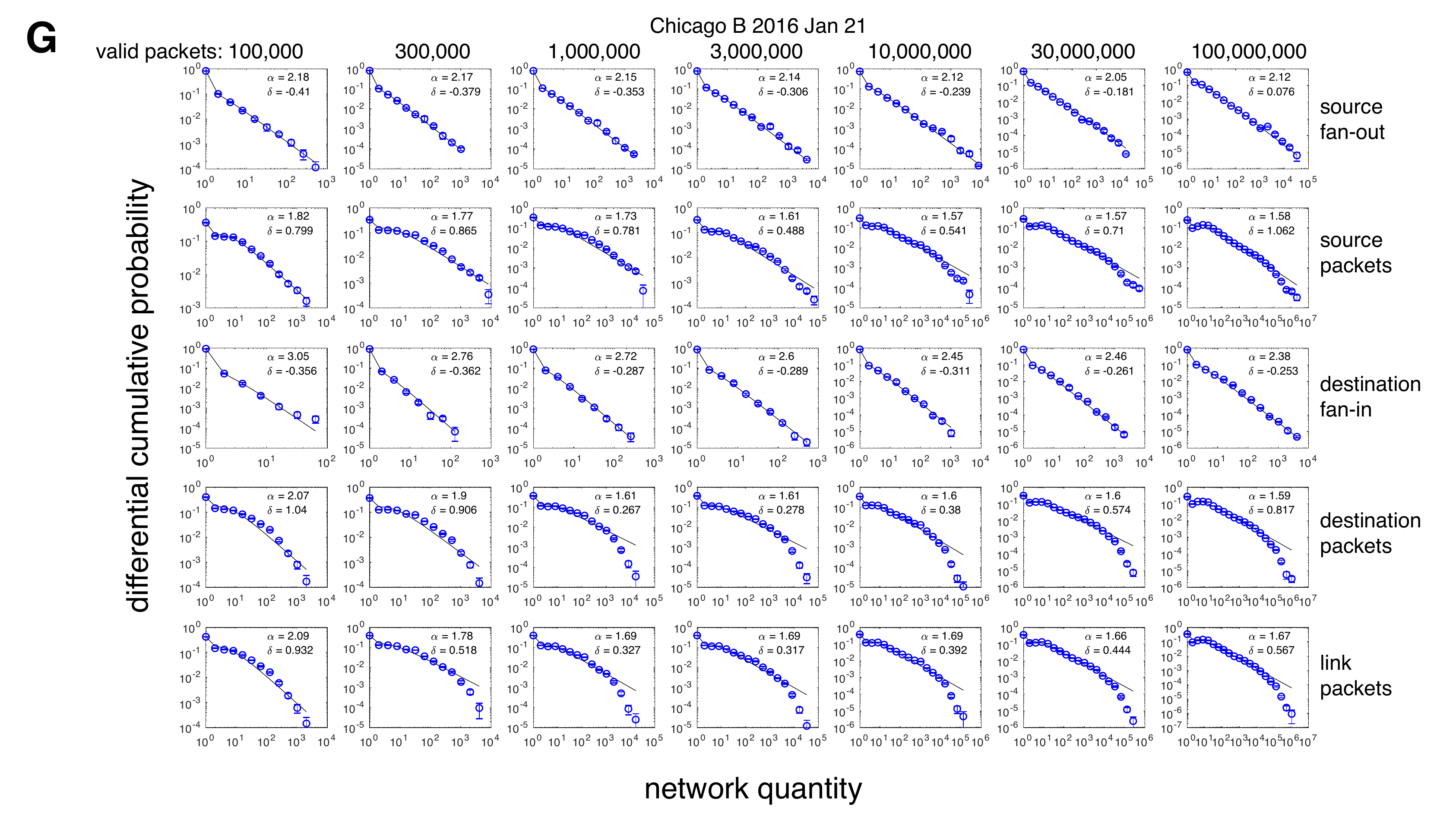}
	\caption{{\bf Measured differential cumulative probabilities.} Blue circles with $\pm$1-$\sigma$ error bars, along with their best-fit modified Zipf--Mandelbrot models (black line) and parameters $\alpha$ and $\delta$ performed for Chicago B 2016 Jan 21.	
	}
	\label{fig:ModelFitsG}
\end{figure}

\begin{figure}[bh]
	\vspace*{-0.5cm}
	\hspace*{-1cm}
	\vspace*{-.9cm}
	\includegraphics[clip, trim=2cm 0cm 0cm 0cm, angle=-90,origin=c,width=\columnwidth]{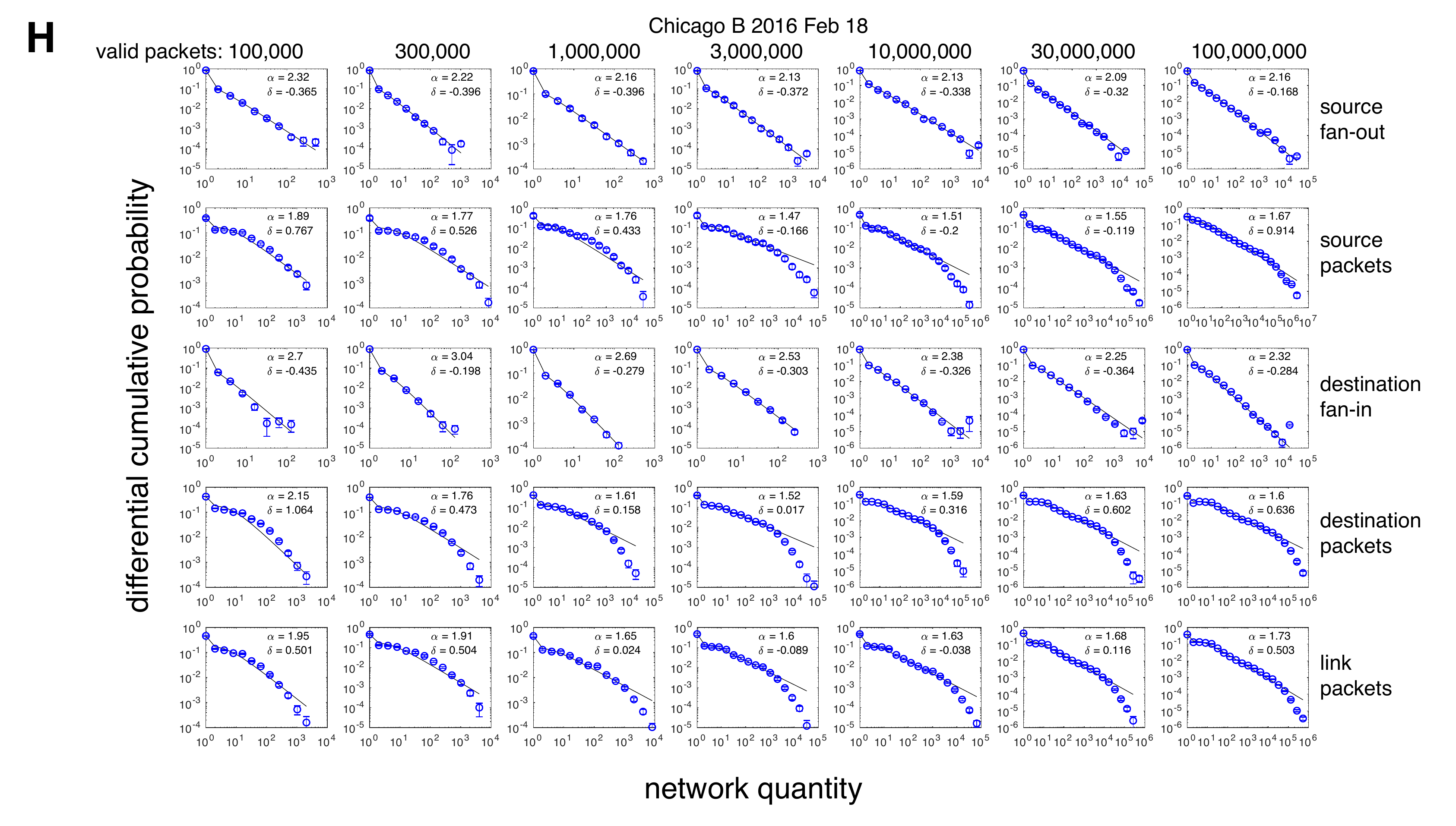}
	\caption{{\bf Measured differential cumulative probabilities.} Blue circles with $\pm$1-$\sigma$ error bars, along with their best-fit modified Zipf--Mandelbrot models (black line) and parameters $\alpha$ and $\delta$ performed for Chicago B 2016 Feb 18.	
	}
	\label{fig:ModelFitsH}
\end{figure}

\begin{figure}[bh]
	\vspace*{-0.5cm}
	\hspace*{-1cm}
	\vspace*{-.9cm}
	\includegraphics[clip, trim=2cm 0cm 0cm 0cm, angle=-90,origin=c,width=\columnwidth]{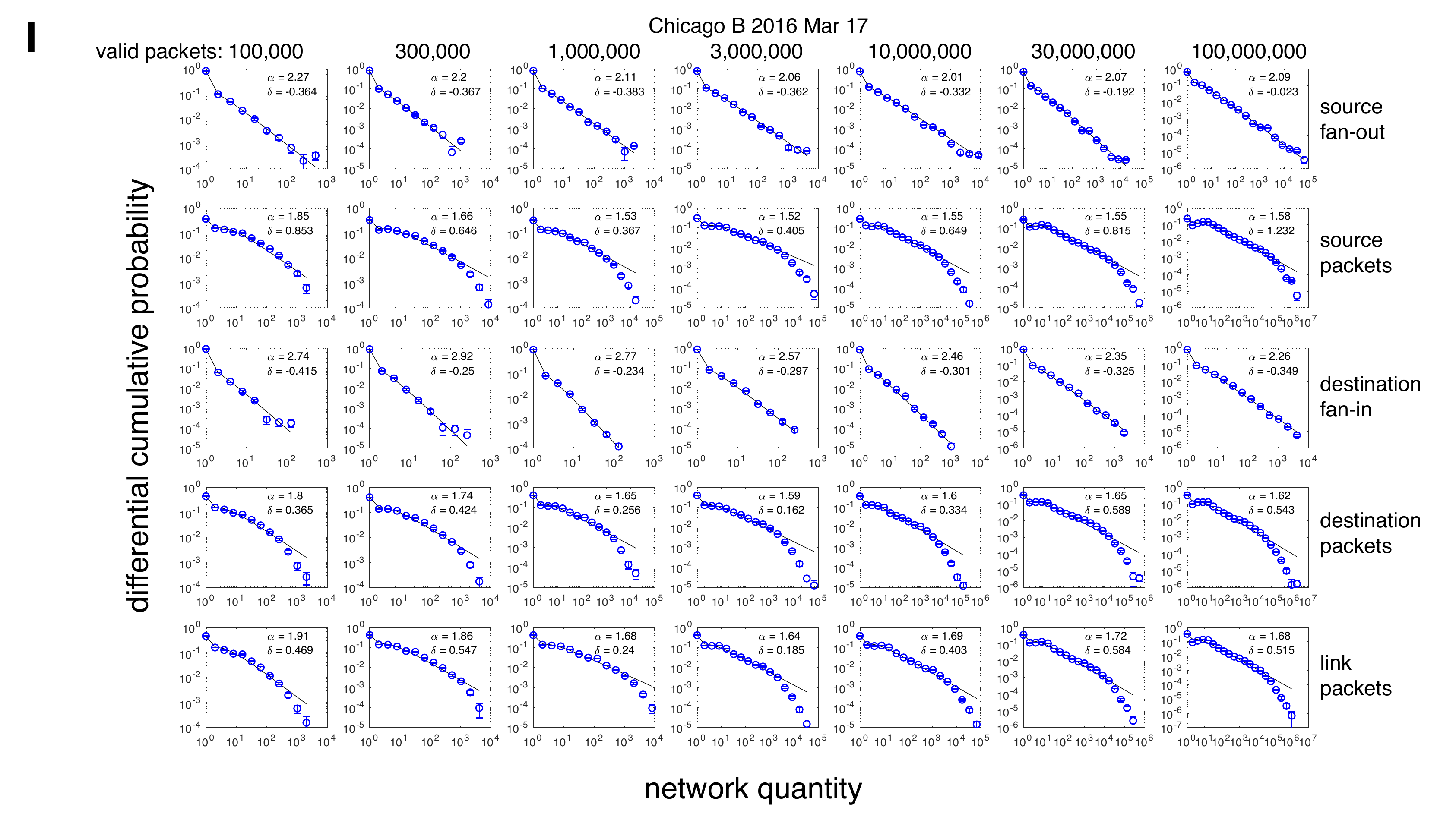}
	\caption{{\bf Measured differential cumulative probabilities.} Blue circles with $\pm$1-$\sigma$ error bars, along with their best-fit modified Zipf--Mandelbrot models (black line) and parameters $\alpha$ and $\delta$ performed for Chicago B 2016 Mar 17.	
	}
	\label{fig:ModelFitsI}
\end{figure}

\begin{figure}[bh]
	\vspace*{-0.5cm}
	\hspace*{-1cm}
	\vspace*{-.9cm}
	\includegraphics[clip, trim=2cm 0cm 0cm 0cm, angle=-90,origin=c,width=\columnwidth]{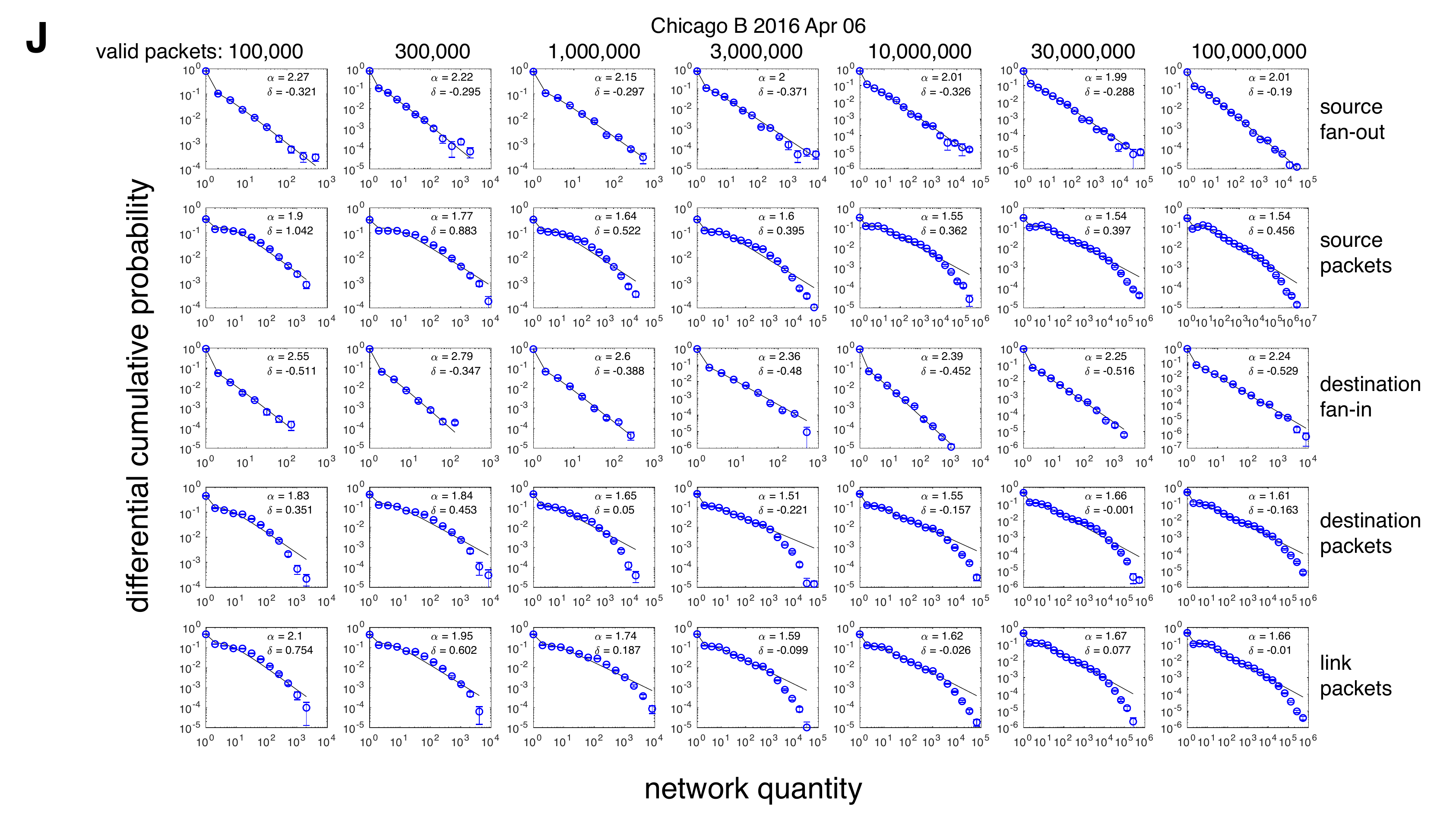}
	\caption{{\bf Measured differential cumulative probabilities.} Blue circles with $\pm$1-$\sigma$ error bars, along with their best-fit modified Zipf--Mandelbrot models (black line) and parameters $\alpha$ and $\delta$ performed for Chicago B 2016 Apr 06.	
	}
	\label{fig:ModelFitsJ}
\end{figure}

\begin{figure}
	\vspace*{-2cm}
	\hspace*{-2cm}
	\includegraphics[width=1.2\columnwidth]{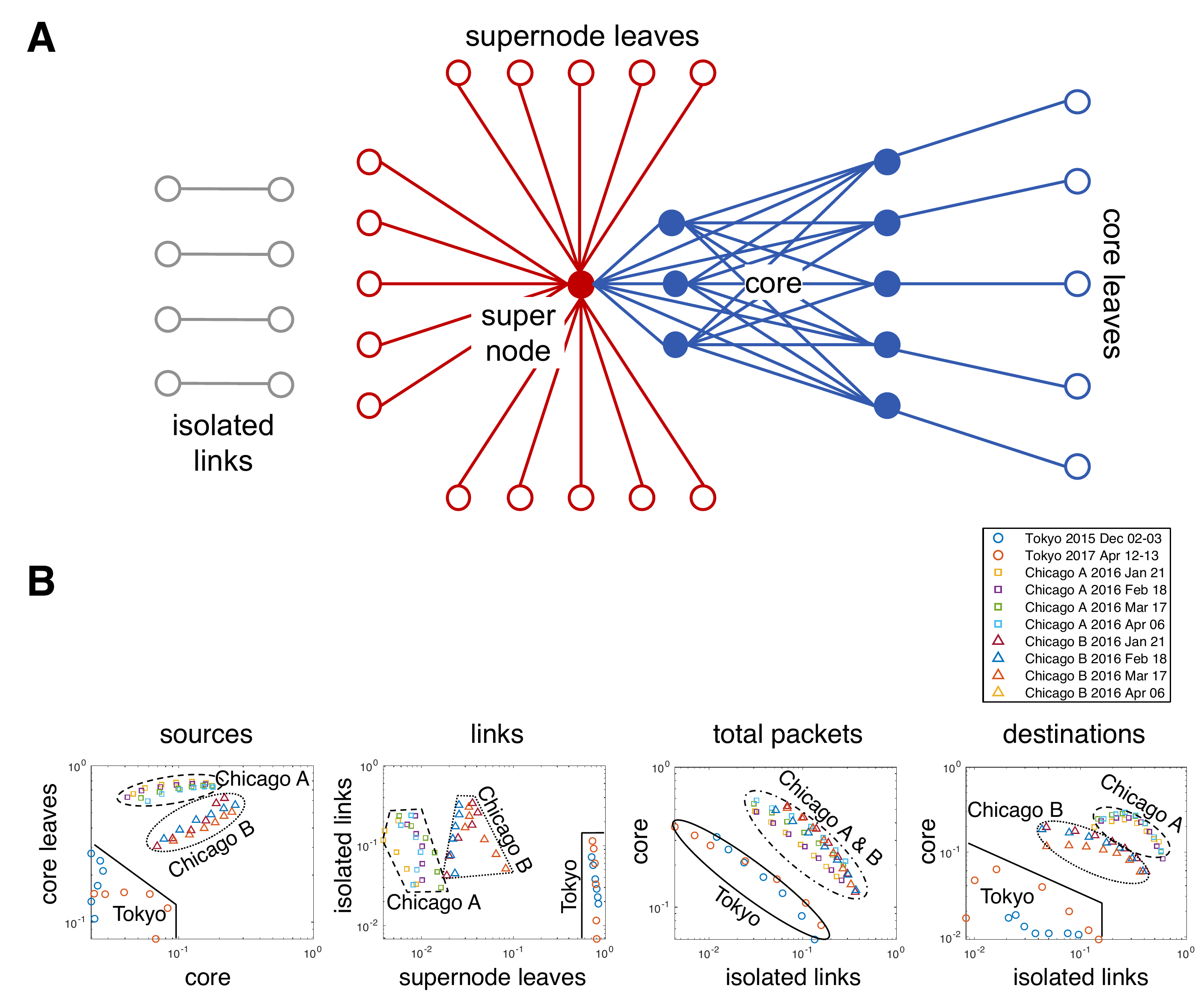}
	\caption{{\bf Distribution of traffic among network topologies.} ({a}) Internet traffic forms networks consisting of a variety of topologies: isolated links, supernode leaves connected to a supernode, and densely connected core(s) with corresponding core leaves. ({b}) A selection of four projections showing the fraction of data in various underlying topologies.  Horizontal and vertical axes are the corresponding fraction of the sources, links, total packets, and destinations that are in various topologies for each location, time, and seven packet windows ($N_V = 10^5, \ldots, 10^8$). These data reveal the differences in the network traffic topologies in the data collected in Tokyo (dominated by supernode leaves), Chicago A (dominated by core leaves), and Chicago B (between Tokyo and Chicago A).}
	\label{fig:NetworkTopology}
\end{figure}

Figure~\ref{fig:NetworkDistribution}b shows five representative model fits out of the 350 performed on 10 data sets, 5 network quantities, and 7 valid packet windows: $N_V = 10^5$, $3{\times}10^5$, $10^6$, $3{\times}10^6$, $10^7$, $3{\times}10^7$, $10^8$.  The model fits are valid over the entire range of $d$ and provide parameter estimates with precisions of 0.01.  In every case, the high value of $p(d=1)$ is indicative of a large contribution from a combination of supernode leaves, core leaves, and isolated links (Figure~\ref{fig:NetworkTopology}a). The breadth and accuracy of these data allow a detailed comparison of the model parameters.  Figure~\ref{fig:NetworkDistribution}c shows the model offset $\delta$ versus the model exponent $\alpha$ for all 350 fits. The different collection locations are clearly distinguishable in this model parameter space.  The Tokyo collections have smaller offsets and are more tightly clustered than the Chicago collections.  Chicago B has a consistently smaller source and link packet model offset than Chicago A.  All the collections have source, link, and destination packet model exponents in the relatively narrow $1.5 < \alpha < 2$ range.  The source fan-out and destination fan-in model exponents are in the broader $1.5 < \alpha < 2.5$ range and are consistent with the prior literature \cite{clauset2009power}.  These results represent an entirely new approach to characterizing Internet traffic that allows the distributions to be projected into a low-dimensional space and enables accurate comparisons among packet collections with different locations, dates, durations, and sizes. Figure~\ref{fig:NetworkDistribution}c indicates that the distributions of the different collection points occupy different parts of the modified Zipf--Mandelbrot model parameter space. Figures~\ref{fig:ModelFitsA}--\ref{fig:ModelFitsJ} show the measured and modeled differential cumulative distributions for the source fan-out, source packets, destination fan-in, destination packets, and link packets for all the collected data.

\begin{figure}[bh]
	\vspace*{-0.5cm}
	\hspace*{-1cm}
	\vspace*{-.2cm}
	\includegraphics[clip, trim=0.9cm 0cm 0cm 0cm, angle=-90,origin=c,width=\columnwidth]{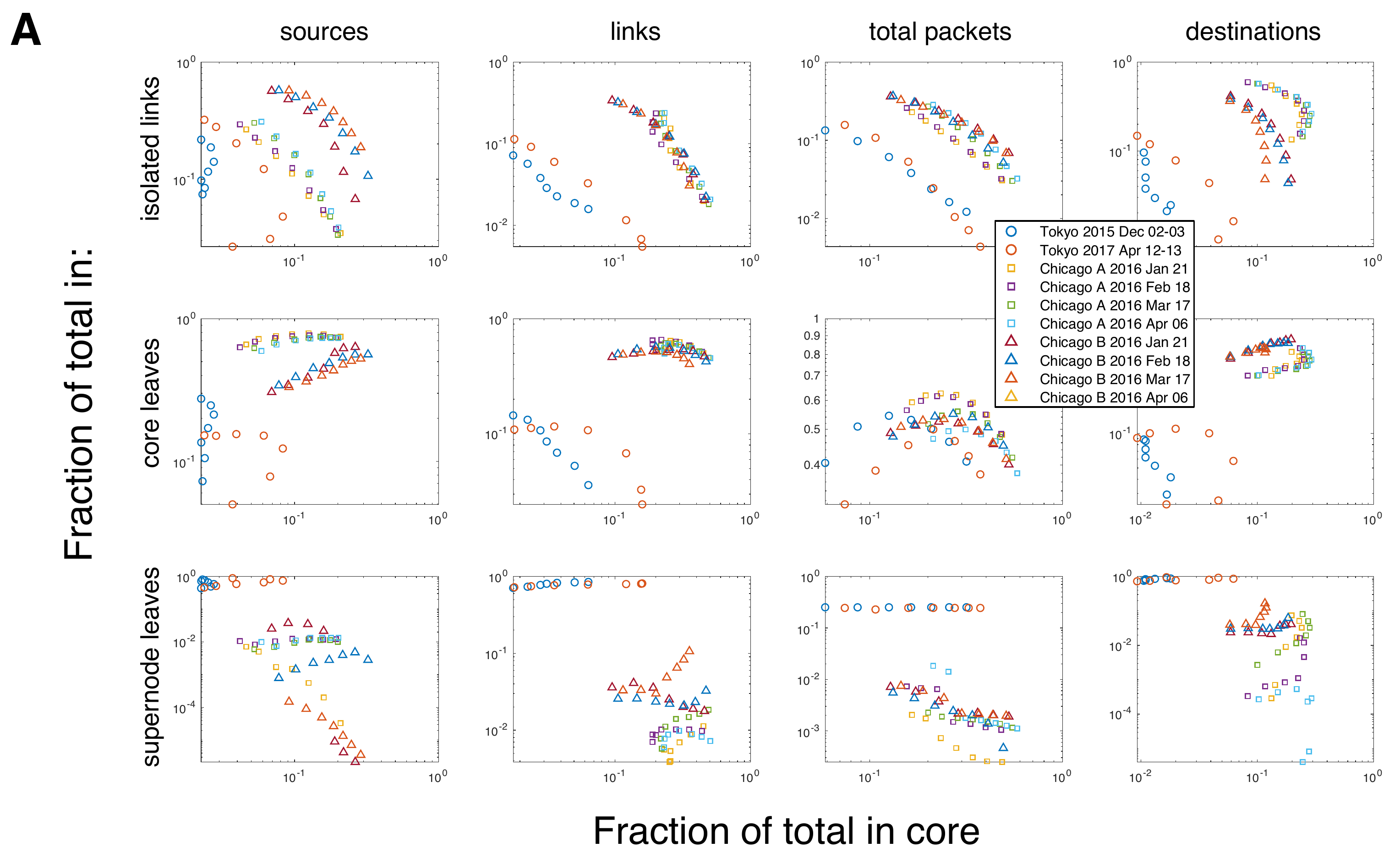}
	\caption{{\bf Fraction of data in different network topologies.}  Fraction of data in isolated links, core leaves, and supernode leaves versus the fraction of data in the core for each location, time, and seven packet windows ($N_V = 10^5, \ldots, 10^8$).
	}
	\label{fig:NetTopoA}
\end{figure}

\begin{figure}[bh]
	\vspace*{-0.5cm}
	\hspace*{-1cm}
	\vspace*{-.2cm}
	\includegraphics[clip, trim=0.9cm 0cm 0cm 0cm, angle=-90,origin=c,width=\columnwidth]{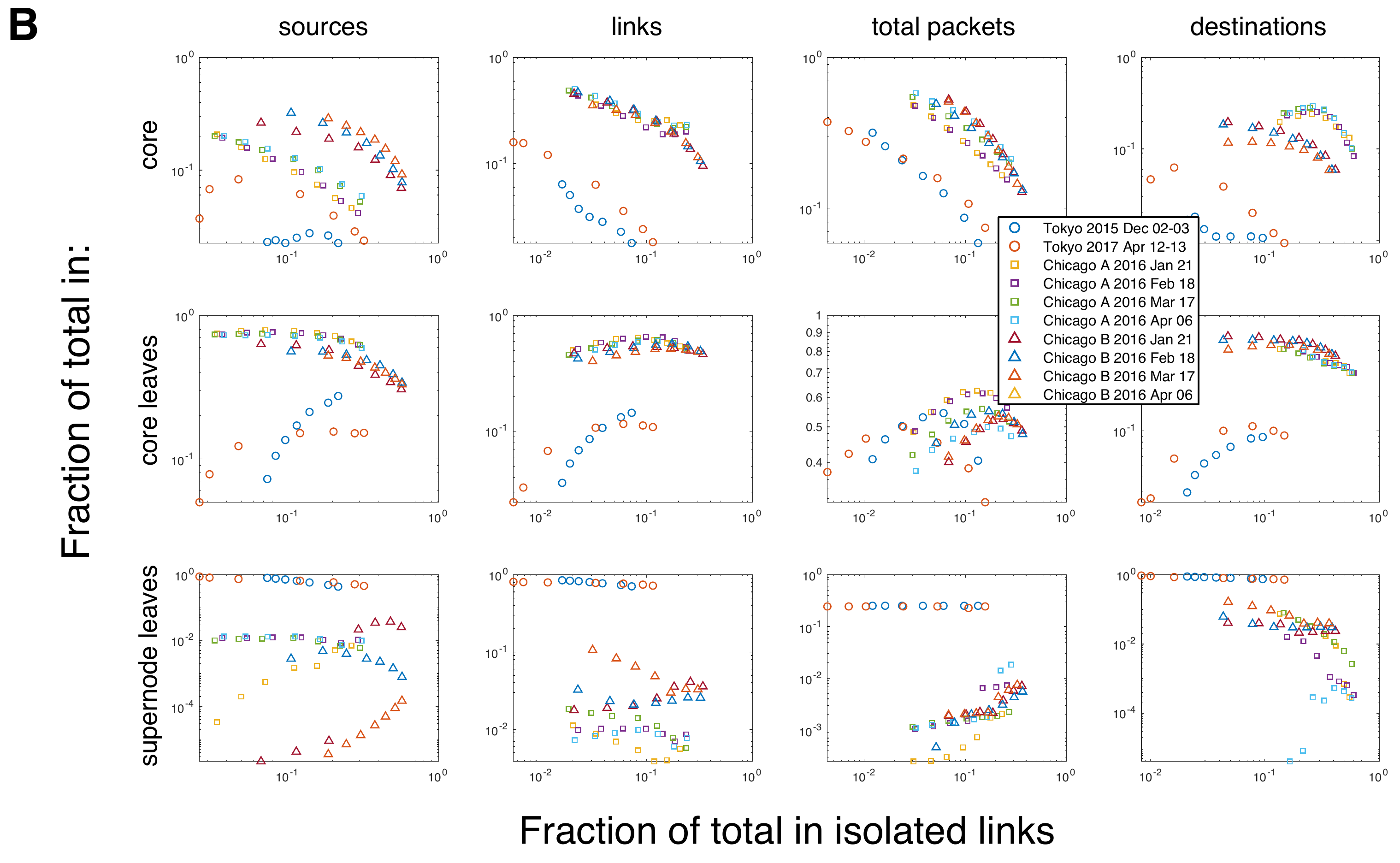}
	\caption{{\bf Fraction of data in different network topologies.}  Fraction of data in the core, core leaves, and supernode leaves versus the fraction of data in isolated links for each location, time, and seven packet windows ($N_V = 10^5, \ldots, 10^8$).
	}
	\label{fig:NetTopoB}
\end{figure}

\begin{figure}[bh]
	\vspace*{-0.5cm}
	\hspace*{-1cm}
	\vspace*{-.2cm}
	\includegraphics[clip, trim=0.9cm 0cm 0cm 0cm, angle=-90,origin=c,width=\columnwidth]{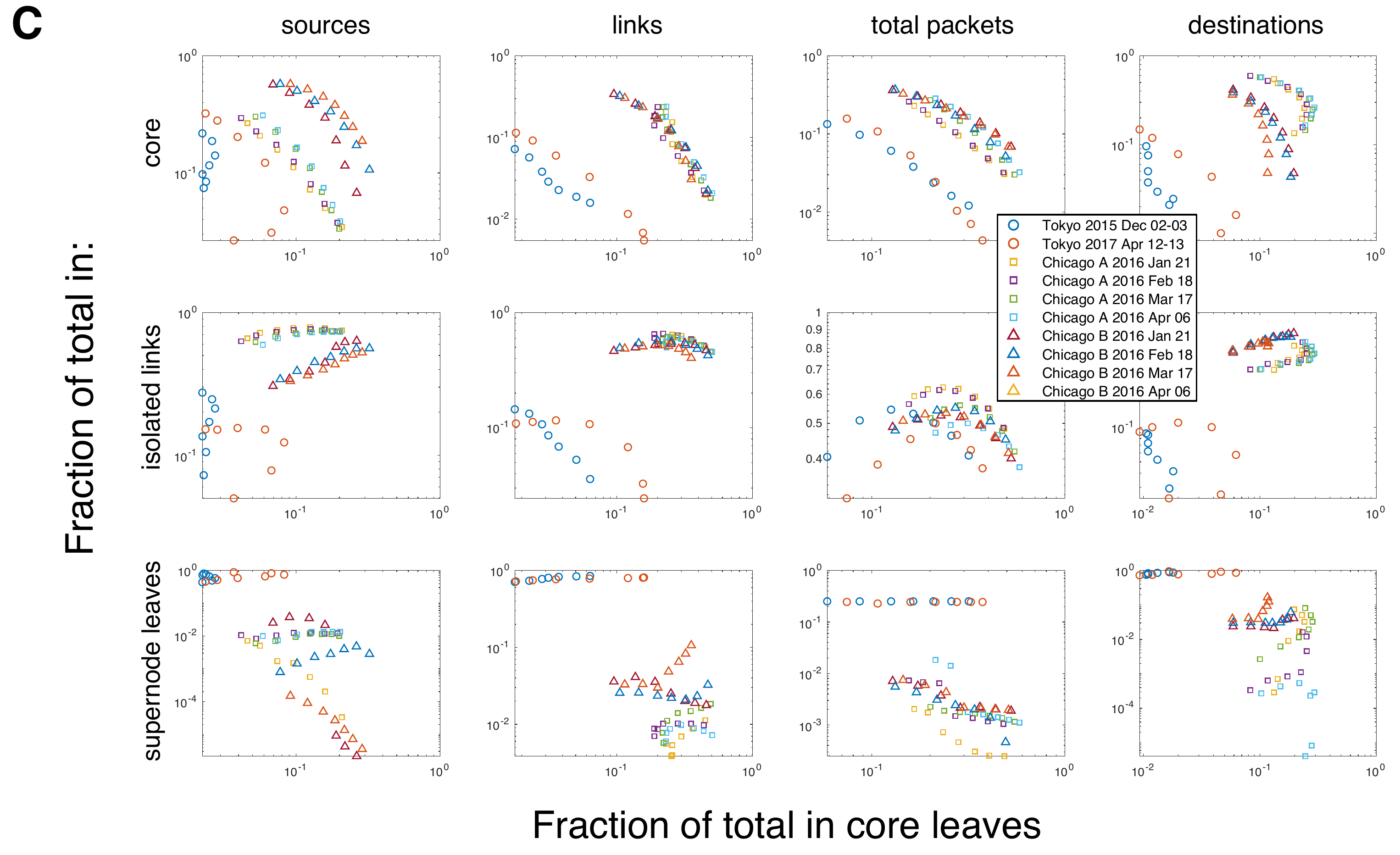}
	\caption{{\bf Fraction of data in different network topologies.}  Fraction of data in the core, isolated links, and supernode leaves versus the fraction of data in core leaves for each location, time, and seven packet windows ($N_V = 10^5, \ldots, 10^8$).
	}
	\label{fig:NetTopC}
\end{figure}

\begin{figure}[bh]
	\vspace*{-0.5cm}
	\hspace*{-1cm}
	\vspace*{-.2cm}
	\includegraphics[clip, trim=0.9cm 0cm 0cm 0cm, angle=-90,origin=c,width=\columnwidth]{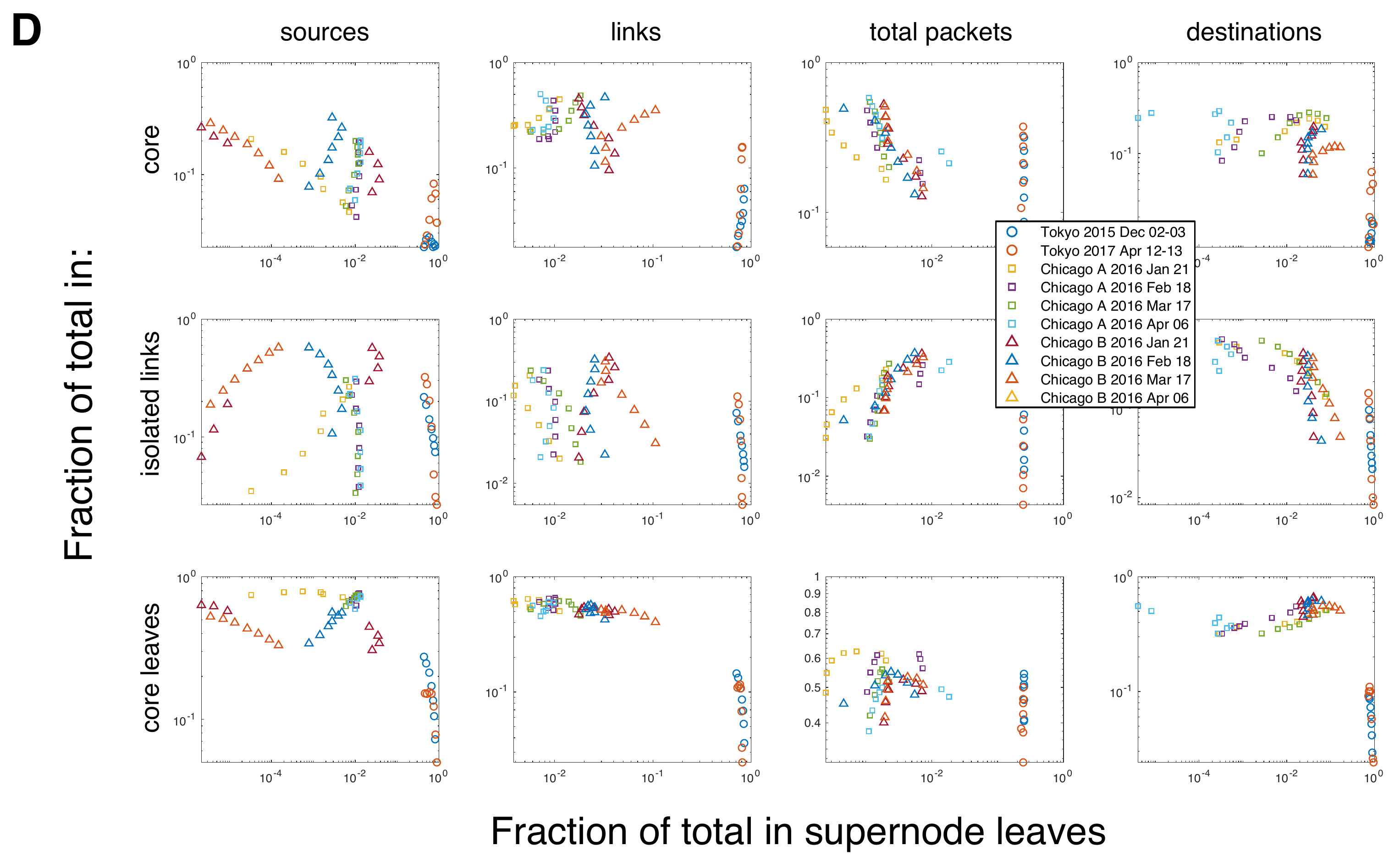}
	\caption{{\bf Fraction of data in different network topologies.}  Fraction of data in the core, isolated links, and core leaves versus the fraction of data in supernode leaves for each location, time, and seven packet windows ($N_V = 10^5, \ldots, 10^8$).
	}
	\label{fig:NetTopD}
\end{figure}

Figure~\ref{fig:NetworkTopology}b shows the average relative fractions of sources, total packets, total links, and the number of destinations in each of the five topologies for the ten data sets, and seven valid packet windows: $N_V = 10^5$, $3{\times}10^5$, $10^6$, $3{\times}10^6$, $10^7$, $3{\times}10^7$, $10^8$.  The four projections in Figure~\ref{fig:NetworkTopology}b are chosen from Figures~\ref{fig:NetTopoA}--\ref{fig:NetTopD} to highlight the differences in the collection locations.  The distinct  regions in the various projections shown in Figure~\ref{fig:NetworkTopology}b  indicate that underlying topological differences are present in the data.  The Tokyo collections have much larger supernode leaf components than the Chicago collections.  The Chicago collections have much larger core and core leaves components than the Tokyo collections.  Chicago A consistently has fewer isolated links than Chicago B.  Comparing the modified Zipf--Mandelbrot model parameters in Figure~\ref{fig:NetworkDistribution}c and underlying topologies in Figure~\ref{fig:NetworkTopology}b suggests that the model parameters are a more compact way to distinguish the network traffic.

\begin{figure}
	\vspace*{-1cm}
	\hspace*{-2cm}
	\includegraphics[width=1.3\columnwidth]{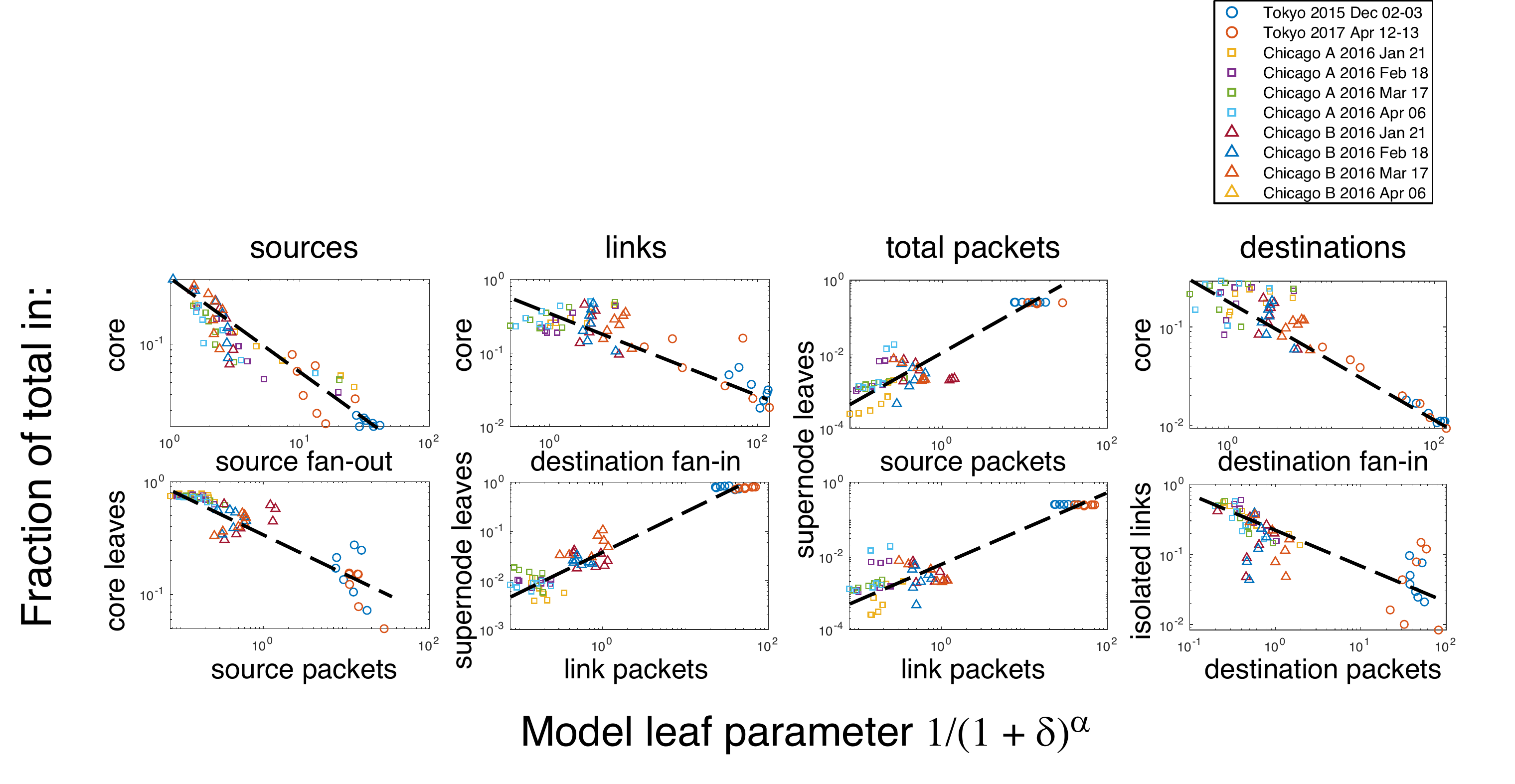}
	\caption{{\bf Topology versus model leaf parameter.} Network topology is highly correlated with the modified Zipf--Mandelbrot model leaf parameter $1/(1+\delta)^\alpha$. A selection of eight projections showing the fraction of data in various underlying topologies.  Vertical axis is the corresponding fraction of the sources, links, total packets, and destinations that are in various topologies. Horizontal axis is the value of the model parameter taken from  either the  source packet, source fan-out, link packet, destination fan-in, or destination packet fits.  Data points are for each location, time, and seven packet windows ($N_V = 10^5, \ldots, 10^8$).}
	\label{fig:TopoModel}
\end{figure}

Figures~\ref{fig:NetworkDistribution}c and \ref{fig:NetworkTopology}b indicate that different collection points  produce different model parameters $\alpha$ and $\delta$ and that these collection points also have different underlying topologies. Figure~\ref{fig:TopoModel} connects the model fits and topology observations by plotting the topology fraction as a function of the model leaf parameter $1/(1+\delta)^\alpha$ which corresponds to the  relative strength of the distribution at $p(d=1)$
\begin{equation}\label{eq:ZipfMandelbrot}
1/(1 + \delta)^\alpha \propto p(d=1;\alpha,\delta)
\end{equation}
The correlations revealed in Figure~\ref{fig:TopoModel} suggest that the model leaf parameter strongly correlates with the fraction of the traffic in different underlying network topologies and is a potentially new and beneficial way to characterize networks.  Figure~\ref{fig:TopoModel} indicates that the fraction of sources, links, and destinations in the core shrinks as the relative importance of the leaf parameter in the source fan-out and destination fan-in increases.  In other words, more source and destination leaves mean a smaller core.    Likewise, the fraction of links and total packets in the supernode leaves grows as the leaf parameter in the link packets and source packets increases.  Interestingly, the fraction of sources in the core leaves and isolated links decreases as the leaf parameter in the source and destination packets increases indicating a shift of sources away from the core leaves and isolated links into supernode leaves.  Thus, the modified Zipf--Mandelbrot model and its leaf parameter provide a direct connection with the network topology, underscoring the value of having accurate model fits across the entire range of values and in particular for $d=1$.

\begin{figure}[bh]
	\vspace*{-0.5cm}
	\includegraphics[clip, trim=1.4cm 0cm 0cm 0cm, angle=-90,origin=c,height=0.75\textheight,width=\columnwidth,  ]{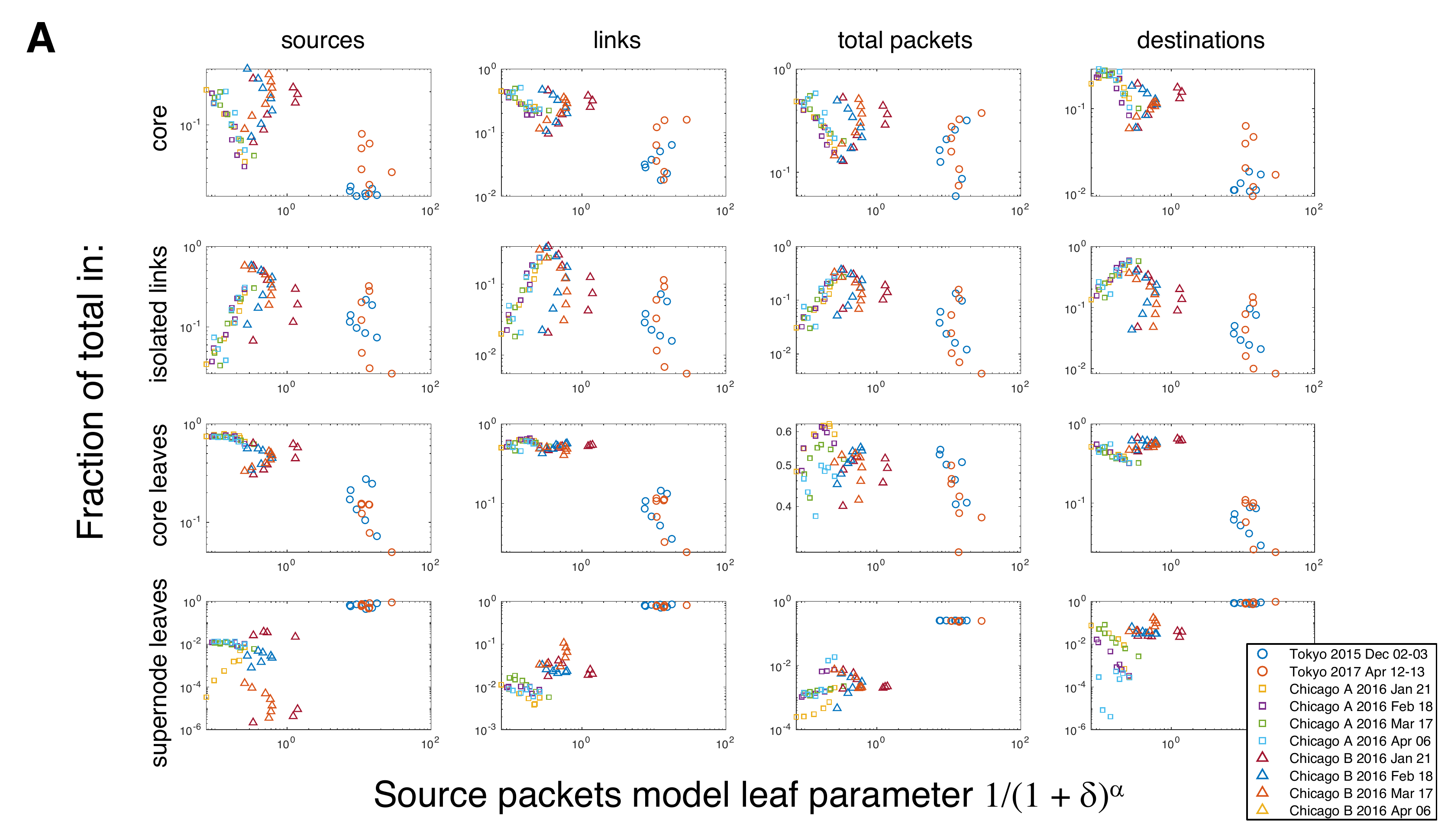}
	\caption{{\bf Topology versus modified Zipf--Mandelbrot model leaf parameter.} Fraction of data in the core, isolated links, core leaves, and supernode leaves versus the model leaf parameter computed from the source packet modified Zipf--Mandelbrot model fits for each location, time, and seven packet windows ($N_V = 10^5, \ldots, 10^8$).
	}
	\label{fig:TopoModelA}
\end{figure}

\begin{figure}[bh]
	\vspace*{-0.5cm}
	\includegraphics[clip, trim=1.4cm 0cm 0cm 0cm, angle=-90,origin=c,height=0.75\textheight,width=\columnwidth,  ]{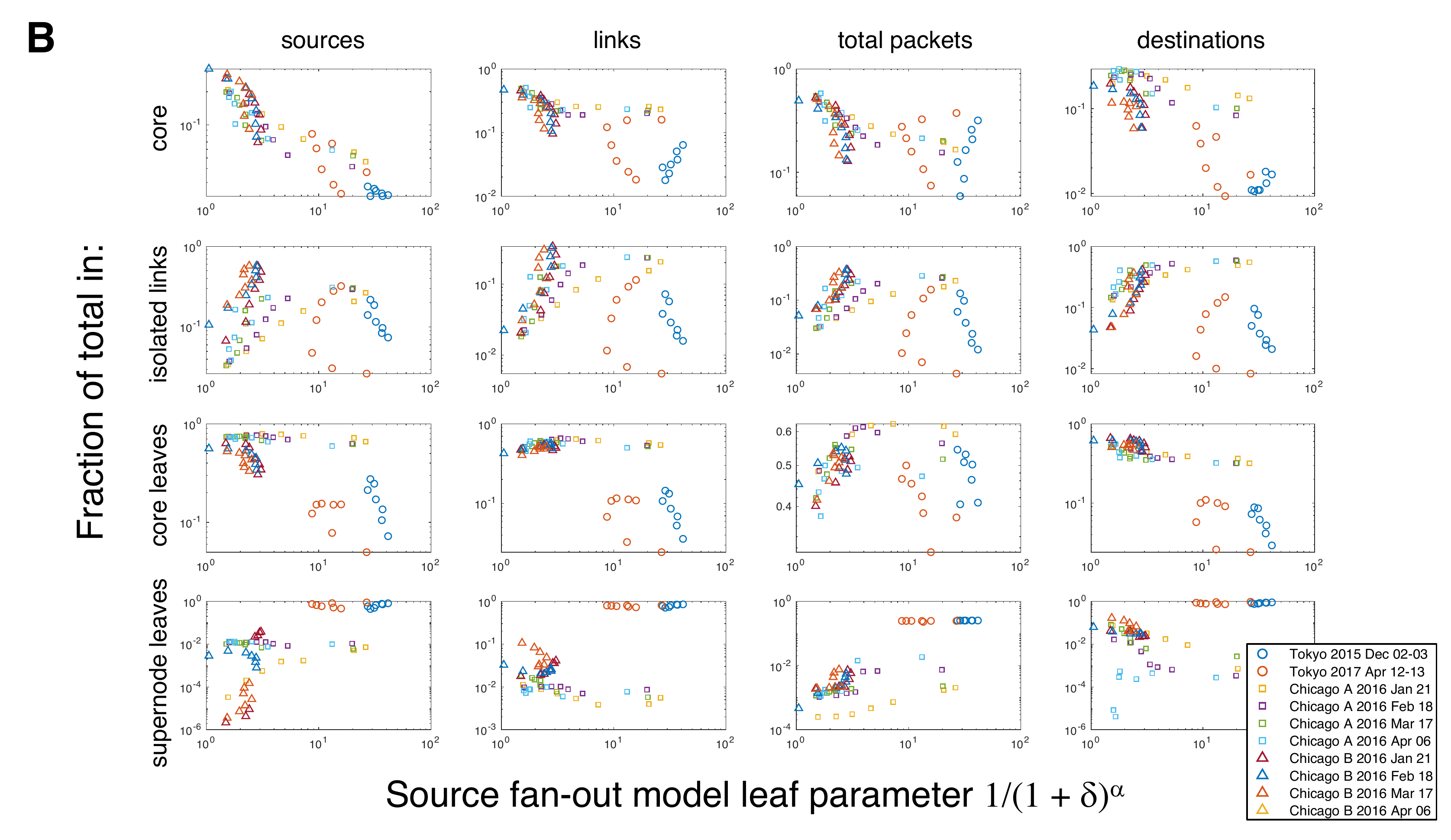}
	\caption{{\bf Topology versus modified Zipf--Mandelbrot model leaf parameter.} Fraction of data in the core, isolated links, core leaves, and supernode leaves versus the model leaf parameter computed from the source fan-out modified Zipf--Mandelbrot model fits for each location, time, and seven packet windows ($N_V = 10^5, \ldots, 10^8$).
	}
	\label{fig:TopoModelB}
\end{figure}

\begin{figure}[bh]
	\vspace*{-0.5cm}
	\includegraphics[clip, trim=1.4cm 0cm 0cm 0cm, angle=-90,origin=c,height=0.75\textheight,width=\columnwidth,  ]{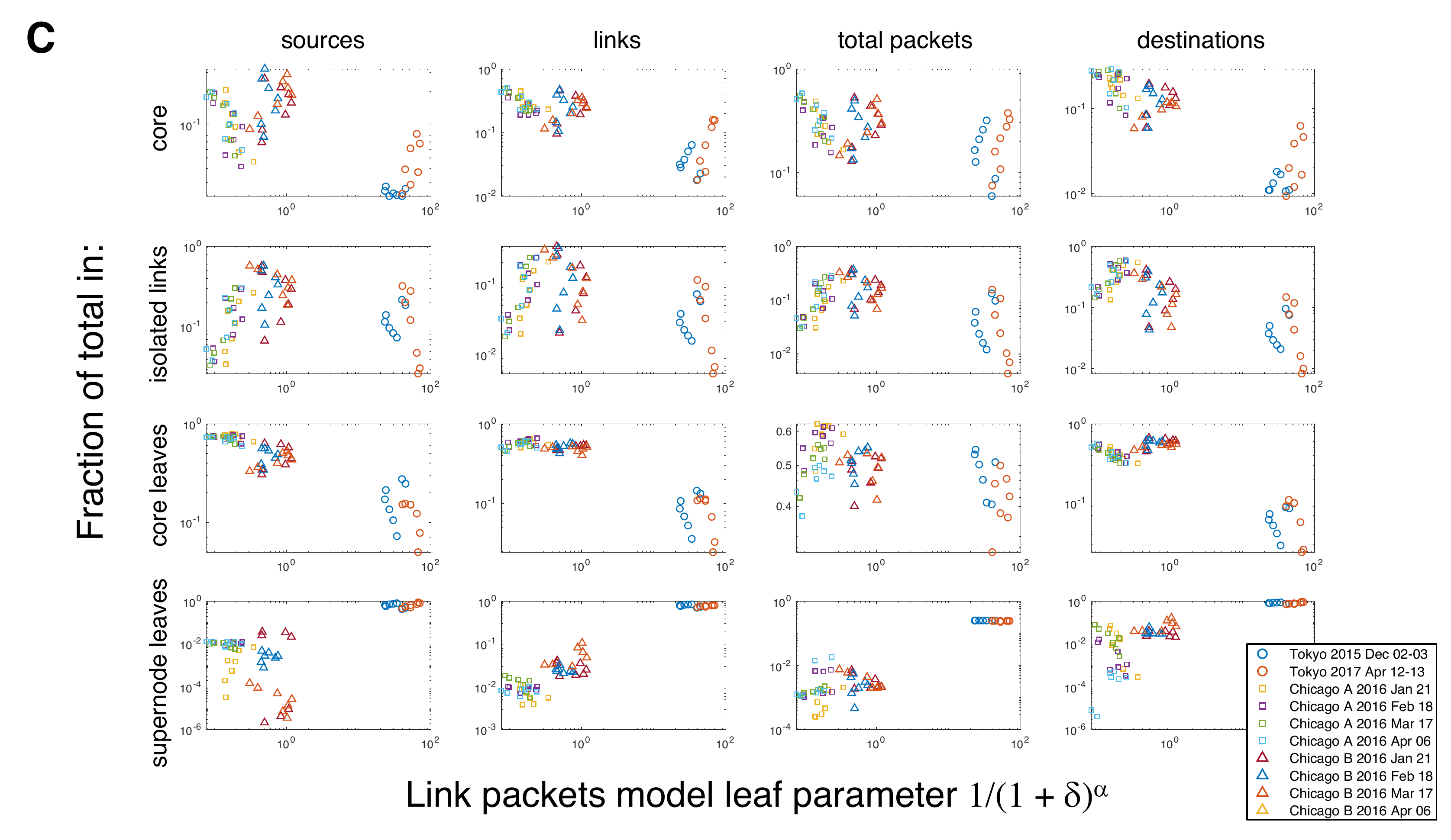}
	\caption{{\bf Topology versus modified Zipf--Mandelbrot model leaf parameter.} Fraction of data in the core, isolated links, core leaves, and supernode leaves versus the model leaf parameter computed from the link packets modified Zipf--Mandelbrot model fits for each location, time, and seven packet windows ($N_V = 10^5, \ldots, 10^8$).
	}
	\label{fig:TopoModelC}
\end{figure}

\begin{figure}[bh]
	\vspace*{-0.5cm}
	\includegraphics[clip, trim=1.4cm 0cm 0cm 0cm, angle=-90,origin=c,height=0.75\textheight,width=\columnwidth,  ]{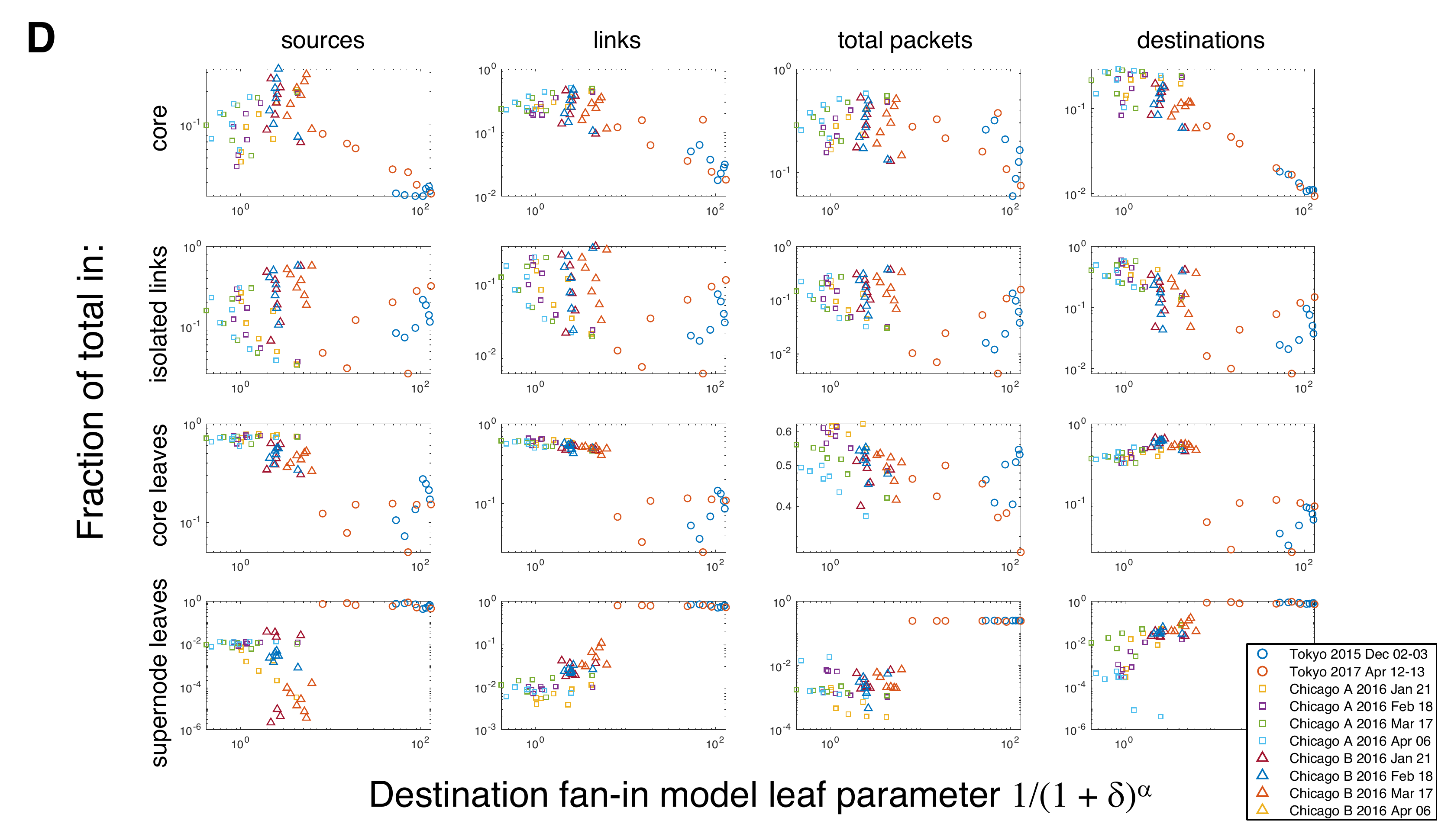}
	\caption{{\bf Topology versus modified Zipf--Mandelbrot model leaf parameter.} Fraction of data in the core, isolated links, core leaves, and supernode leaves versus the model leaf parameter computed from the destination fan-out modified Zipf--Mandelbrot model fits for each location, time, and seven packet windows ($N_V = 10^5, \ldots, 10^8$).
	}
	\label{fig:TopoModelD}
\end{figure}

\begin{figure}[bh]
	\vspace*{-0.5cm}
	\includegraphics[clip, trim=1.4cm 0cm 0cm 0cm, angle=-90,origin=c,height=0.75\textheight,width=\columnwidth,  ]{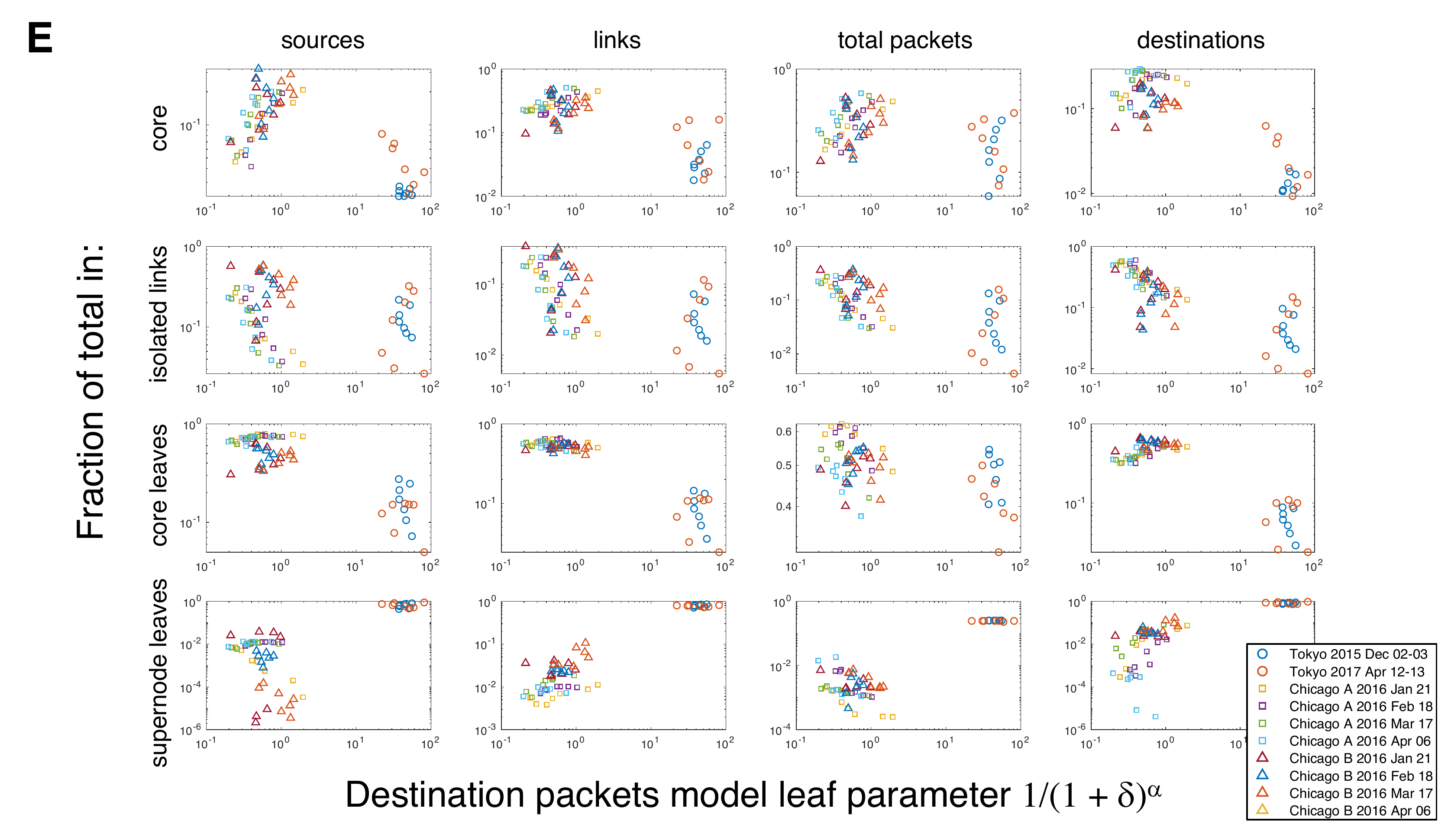}
	\caption{{\bf Topology versus modified Zipf--Mandelbrot model leaf parameter.} Fraction of data in the core, isolated links, core leaves, and supernode leaves versus the model leaf parameter computed from the destination packets modified Zipf--Mandelbrot model fits for each location, time, and seven packet windows ($N_V = 10^5, \ldots, 10^8$).
	}
	\label{fig:TopoModelE}
\end{figure}

Figures~\ref{fig:TopoModelA}--\ref{fig:TopoModelE} show the fraction of the sources, links, total packets, and destinations in each of the measured topologies for all the locations as a function of the modified Zipf--Mandelbrot leaf parameter computed from the model fits of the source packets, source fan-out, link packets, destination fan-in, and destination packets taken from Figures~\ref{fig:ModelFitsA}--\ref{fig:ModelFitsJ}.

\section{Discussion}\label{Discussion}
Measurements of Internet traffic are useful for informing policy, identifying and preventing outages, defeating attacks, planning for future loads, and protecting the Domain Name System \cite{clark20179th}.
On a given day, millions of IPs are engaged in scanning behavior. Our improved models can aid cybersecurity analysts in determining which of these IPs are nefarious \cite{yu2012predicted}, the distribution of attacks in particular critical sectors \cite{husak2018assessing}, identifying spamming behavior \cite{fonseca2016measuring}, how to vaccinate against computer viruses \cite{balthrop2004technological}, obscuring web sources \cite{javed2015measurement}, identifying significant flow aggregates in traffic \cite{cho2017recursive}, and sources of rumors \cite{paluch2018fast}.

The results presented here have a number of potential practical applications for Internet stakeholders.  The methods presented of collecting, filtering, computing, and binning the data to produce accurate measurements of a variety of network quantities are generally applicable to Internet measurements and have the potential to produce more accurate measures of these quantities.  The accurate fits of the two-parameter modified Zipf--Mandelbrot distribution offer all the usual benefits of low-parameter models: measuring parameters with far less data, accurate predictions of network quantities based on a few parameters, observing changes in the underlying distribution, and using modeled distributions to detect anomalies in the data.

From a scientific perspective, improved knowledge of how Internet traffic flows can inform our understanding of how economics, topology, and demand shape the Internet over time.  As with all scientific disciplines, the ability of theoreticians to develop and test theories of the Internet and network phenomena is bounded by the scale and accuracy of measured phenomena \cite{adamic2000power,bohman2009emergence,stumpf2012critical,virkar2014power}. The connections among dynamic evolution \cite{bianconi2001competition}, network topology \cite{mucha2010community}\cite{boccaletti2014structure,lu2016vital}, network robustness \cite{li2017fundamental}, controllability \cite{liu2016control}, community formation \cite{perc2017statistical}, and spreading phenomena \cite{holme2015modern} have emerged in many contexts \cite{barabasi2009scale,wang2016statistical,koliba2018governance}.  Many first-principles theories for Internet and network phenomena have been proposed, such as Poisson models \cite{paxson1995wide}, fractional Brownian motion \cite{willinger1997self}, preferential attachment \cite{barabasi1999emergence,albert1999internet}\cite{newman2001clustering,sheridan2018preferential}, statistical mechanics \cite{albert2002statistical}, percolation \cite{achlioptas2009explosive}, hyperbolic geometries \cite{krioukov2009curvature,krioukov2010hyperbolic}, non-global greedy routing \cite{boguna2009navigating,boguna2009navigability,boguna2010sustaining}, interacting particle systems \cite{antonopoulos2018opinion}, higher-order organization of complex networks from graph motifs \cite{benson2016higher}, and minimum control energy \cite{lindmark2018minimum}.  All of these models require data to test them. In contrast to previous network models that have principally been based on data obtained from network crawls from a variety of start points on the network, our network traffic data are collected from observations of network streams.  Both viewpoints provide important network observations. Observations of a network stream provide complementary data on network dynamics and highlight the contribution of leaves and isolated edges, which are less sampled in network crawls.

The aggregated data sets our teams have collected provide a unique window into these questions.  The nonlinear fitting techniques described are a novel approach to fitting power-law data and have potential applications to power-law networks in diverse domains. The model fit parameters present new opportunities to connect the distributions to underlying theoretical models of networks. That the model fit parameters distinguish the different collection points and are reflective of different network topologies in the data at these points suggests a deeper underlying connection between the models and the network topologies.

\section{Conclusions}\label{Conclusions}

Our society critically depends on the Internet for our professional, personal, and political lives. This dependence has rapidly grown much stronger than our comprehension of its underlying structure, performance limits, dynamics, and evolution. The fundamental characteristics of the Internet are perpetually challenging to research and analyze, and we must admit we know little about what keeps the system stable. As a result, researchers and policymakers deal with a multi-trillion-dollar ecosystem essentially in the dark, and agencies charged with infrastructure protection have little situational awareness regarding global dynamics and operational threats.  This paper has presented an analysis of the largest publicly available collection of Internet traffic consisting of 50 billion packets and reveals a new phenomenon: the importance of otherwise unseen leaf nodes and isolated links in Internet traffic.  Our analysis further shows that a two-parameter modified Zipf--Mandelbrot distribution accurately describes a wide variety of source/destination statistics on moving sample windows ranging from 100{,}000 to 100{,}000{,}000 packets over collections that span years and continents. The measured model parameters distinguish different network streams, and the model leaf parameter strongly correlates with the fraction of the traffic in different underlying network topologies.  These results represent a significant improvement in Internet modeling accuracy, improve our understanding of the Internet, and show the importance of stream sampling for measuring network phenomena.

\let\cleardoublepage\clearpage
\section*{Acknowledgements}
The authors wish to acknowledge the following individuals for their contributions and support: Shohei  Araki, William Arcand, David Bestor, William Bergeron, Bob Bond, Paul Burkhardt, Chansup Byun, Cary Conrad, Alan Edelman, Sterling Foster, Bo Hu, Matthew Hubbell, Micheal Houle, Micheal Jones, Anne Klein, Charles Leiserson, Dave Martinez, Mimi McClure, Julie Mullen, Steve Pritchard, Andrew Prout, Albert Reuther, Antonio Rosa, Victor Roytburd, Siddharth Samsi, Koichi Suzuki, Kenji Takahashi, Michael Wright, Charles Yee, and Michitoshi  Yoshida.


\printindex
\cleardoublepage
\end{document}

%% file: cm-macro.tex
\usepackage{xcolor,marginnote}
\definecolor{gry}{cmyk}{0,0,0,.2}
\def\AQ#1#2#3{\bgroup\ignorespaces{\fboxsep0pt\colorbox{gry}{#1}}\marginpar{\vskip#2\fboxsep4pt\colorbox{gry}{\begin{minipage}{3.5pc}\raggedright\footnotesize #3\end{minipage}}}\ignorespaces\egroup}%

\def\onlyfordigital#1{\iffalse #1\fi}

%% file: Run_Chapter.bbl
\begin{thebibliography}{100}

\bibitem{achlioptas2009explosive}
Dimitris Achlioptas, Raissa~M. D'souza, and Joel Spencer.
\newblock Explosive percolation in random networks.
\newblock {\em Science}, 323(5920):1453--1455, 2009.

\bibitem{adamic2000power}
Lada~A. Adamic and Bernardo~A. Huberman.
\newblock Power-law distribution of the world wide web.
\newblock {\em Science}, 287(5461):2115--2115, 2000.

\bibitem{albert2002statistical}
R{\'e}ka Albert and Albert-L{\'a}szl{\'o} Barab{\'a}si.
\newblock Statistical mechanics of complex networks.
\newblock {\em Reviews of Modern Physics}, 74(1):47, 2002.

\bibitem{albert1999internet}
R{\'e}ka Albert, Hawoong Jeong, and Albert-L{\'a}szl{\'o} Barab{\'a}si.
\newblock Internet: Diameter of the world-wide web.
\newblock {\em Nature}, 401(6749):130, 1999.

\bibitem{allman2007brief}
Mark Allman, Vern Paxson, and Jeff Terrell.
\newblock A brief history of scanning.
\newblock In {\em Proceedings of the 7th ACM SIGCOMM Conference on Internet
  measurement}, pages 77--82. ACM, 2007.

\bibitem{antonopoulos2018opinion}
Chris~G. Antonopoulos and Yilun Shang.
\newblock Opinion formation in multiplex networks with general initial
  distributions.
\newblock {\em Scientific Reports}, 8(1):2852, 2018.

\bibitem{bader2013graph}
David~A. Bader, Henning Meyerhenke, Peter Sanders, and Dorothea Wagner.
\newblock {\em Graph Partitioning and Graph Clustering}, volume 588.
\newblock American Mathematical Society, Ann Arbor, 2013.

\bibitem{balthrop2004technological}
Justin Balthrop, Stephanie Forrest, Mark~E.J. Newman, and Matthew~M. Williamson.
\newblock Technological networks and the spread of computer viruses.
\newblock {\em Science}, 304(5670):527--529, 2004.

\bibitem{barabasi2009scale}
Albert-L{\'a}szl{\'o} Barab{\'a}si.
\newblock Scale-free networks: A decade and beyond.
\newblock {\em Science}, 325(5939):412--413, 2009.

\bibitem{barabasi1999emergence}
Albert-L{\'a}szl{\'o} Barab{\'a}si and R{\'e}ka Albert.
\newblock Emergence of scaling in random networks.
\newblock {\em Science}, 286(5439):509--512, 1999.

\bibitem{barabasi2016network}
Albert-L{\'a}szl{\'o} Barab{\'a}si et~al.
\newblock {\em Network Science}.
\newblock Cambridge University Press, Cambridge, 2016.

\bibitem{benson2016higher}
Austin~R. Benson, David~F. Gleich, and Jure Leskovec.
\newblock Higher-order organization of complex networks.
\newblock {\em Science}, 353(6295):163--166, 2016.

\bibitem{bharti2010inferring}
Vineet Bharti, Pankaj Kankar, Lokesh Setia, Gonca G{\"u}rsun, Anukool Lakhina,
  and Mark Crovella.
\newblock Inferring invisible traffic.
\newblock In {\em Proceedings of the 6th International Conference}, page~22.
  ACM, 2010.

\bibitem{bianconi2001competition}
Ginestra Bianconi and A.-L. Barab{\'a}si.
\newblock Competition and multiscaling in evolving networks.
\newblock {\em EPL (Europhysics Letters)}, 54(4):436, 2001.

\bibitem{boccaletti2014structure}
Stefano Boccaletti, Ginestra Bianconi, Regino Criado, Charo~I. Del~Genio,
  Jes{\'u}s G{\'o}mez-Gardenes, Miguel Romance, Irene Sendina-Nadal, Zhen Wang,
  and Massimiliano Zanin.
\newblock The structure and dynamics of multilayer networks.
\newblock {\em Physics Reports}, 544(1):1--122, 2014.

\bibitem{boguna2009navigating}
Mari{\'a}n Bogun{\'a} and Dmitri Krioukov.
\newblock Navigating ultrasmall worlds in ultrashort time.
\newblock {\em Physical Review Letters}, 102(5):058701, 2009.

\bibitem{boguna2009navigability}
Marian Boguna, Dmitri Krioukov, and Kimberly~C. Claffy.
\newblock Navigability of complex networks.
\newblock {\em Nature Physics}, 5(1):74, 2009.

\bibitem{boguna2010sustaining}
Mari{\'a}n Bogun{\'a}, Fragkiskos Papadopoulos, and Dmitri Krioukov.
\newblock Sustaining the internet with hyperbolic mapping.
\newblock {\em Nature Communications}, 1:62, 2010.

\bibitem{bohman2009emergence}
Tom Bohman.
\newblock Emergence of connectivity in networks.
\newblock {\em Evolution}, 11:13, 2009.

\bibitem{borgnat2009seven}
Pierre Borgnat, Guillaume Dewaele, Kensuke Fukuda, Patrice Abry, and Kenjiro
  Cho.
\newblock Seven years and one day: Sketching the evolution of internet traffic.
\newblock In {\em INFOCOM 2009, IEEE}, pages 711--719. IEEE, 2009.

\bibitem{Brbic2018}
Maria~Brbic and  Ivica~Kopriva.
\newblock $\ell_0$-motivated low-rank sparse subspace clustering.
\newblock {\em IEEE Transactions on Cybernetics}, 50:1--15, 2018.

\bibitem{broder2000graph}
Andrei Broder, Ravi Kumar, Farzin Maghoul, Prabhakar Raghavan, Sridhar
  Rajagopalan, Raymie Stata, Andrew Tomkins, and Janet Wiener.
\newblock Graph structure in the web.
\newblock {\em Computer Networks}, 33(1-6):309--320, 2000.

\bibitem{cao2009identifying}
Jing Cao, Yu~Jin, Aiyou Chen, Tian Bu, and Z-L Zhang.
\newblock Identifying high cardinality internet hosts.
\newblock In {\em INFOCOM 2009, IEEE}, pages 810--818. IEEE, 2009.

\bibitem{chartrand2007exact}
Rick Chartrand.
\newblock Exact reconstruction of sparse signals via nonconvex minimization.
\newblock {\em IEEE Signal Processing Letters}, 14(10):707--710, 2007.

\bibitem{chiu2015we}
Yi-Ching Chiu, Brandon Schlinker, Abhishek~Balaji Radhakrishnan, Ethan
  Katz-Bassett, and Ramesh Govindan.
\newblock Are we one hop away from a better internet?
\newblock In {\em Proceedings of the 2015 Internet Measurement Conference},
  pages 523--529. ACM, 2015.

\bibitem{cho2017recursive}
Kenjiro Cho.
\newblock Recursive lattice search: Hierarchical heavy hitters revisited.
\newblock In {\em Proceedings of the 2017 Internet Measurement Conference},
  pages 283--289. ACM, 2017.

\bibitem{cho2006impact}
Kenjiro Cho, Kensuke Fukuda, Hiroshi Esaki, and Akira Kato.
\newblock The impact and implications of the growth in residential user-to-user
  traffic.
\newblock In {\em ACM SIGCOMM Computer Communication Review}, volume~36, pages
  207--218. ACM, 2006.

\bibitem{cho2008observing}
Kenjiro Cho, Kensuke Fukuda, Hiroshi Esaki, and Akira Kato.
\newblock Observing slow crustal movement in residential user traffic.
\newblock In {\em Proceedings of the 2008 ACM CoNEXT Conference}, page~12. ACM,
  2008.

\bibitem{cho2000tr}
Kenjiro Cho, Koushirou Mitsuya, and Akira Kato.
\newblock Traffic data repository at the wide project.
\newblock In {\em Proceedings of USENIX 2000 Annual Technical Conference:
  FREENIX Track}, pages 263--270, 2000.

\bibitem{claffy1999internet}
Kimberly~C.~Claffy.
\newblock Internet tomography.
\newblock {\em Nature, Web Matter}, 1999.

\bibitem{claffy2000measuring}
Kimberly~C.~Claffy.
\newblock Measuring the internet.
\newblock {\em IEEE Internet Computing}, 4(1):73--75, 2000.

\bibitem{claffy1994tracking}
Kimberly~C. Claffy, Hans-Werner Braun, and George~C. Polyzos.
\newblock Tracking long-term growth of the nsfnet.
\newblock {\em Communications of the ACM}, 37(8):34--45, 1994.

\bibitem{clark20179th}
David Clark et~al.
\newblock The 9th workshop on active internet measurements (aims-9) report.
\newblock {\em ACM SIGCOMM Computer Communication Review}, 47(5):35--38, 2017.

\bibitem{clauset2009power}
Aaron Clauset, Cosma~Rohilla Shalizi, and Mark~E.J. Newman.
\newblock Power-law distributions in empirical data.
\newblock {\em SIAM Review}, 51(4):661--703, 2009.

\bibitem{dainotti2012issues}
Alberto Dainotti, Antonio Pescape, and Kimberly~C. Claffy.
\newblock Issues and future directions in traffic classification.
\newblock {\em IEEE Network}, 26(1), 2012.

\bibitem{dhamdhere2018inferring}
Amogh Dhamdhere, David~D. Clark, Alexander Gamero-Garrido, Matthew Luckie,
  Ricky~K.P. Mok, Gautam Akiwate, Kabir Gogia, Vaibhav Bajpai, Alex~C. Snoeren,
  and Kimberly~C.~Claffy.
\newblock Inferring persistent interdomain congestion.
\newblock In {\em Proceedings of the 2018 Conference of the ACM Special
  Interest Group on Data Communication}, pages 1--15. ACM, 2018.

\bibitem{dhamdhere2010internet}
Amogh Dhamdhere and Constantine Dovrolis.
\newblock The internet is flat: modeling the transition from a transit
  hierarchy to a peering mesh.
\newblock In {\em Proceedings of the 6th International Conference}, page~21.
  ACM, 2010.

\bibitem{NIST:DLMF}
{\it NIST Digital Library of Mathematical Functions, Release 1.0.20 of
  2018-09-15}.
\newblock http://dlmf.nist.gov/25.11.
\newblock F.~W.~J. Olver, A.~B. {Olde Daalhuis}, D.~W. Lozier, B.~I. Schneider,
  R.~F. Boisvert, C.~W. Clark, B.~R. Miller and B.~V. Saunders, eds.

\bibitem{donoho2006compressed}
David~L. Donoho.
\newblock Compressed sensing.
\newblock {\em IEEE Transactions on Information Theory}, 52(4):1289--1306,
  2006.

\bibitem{dovrolis2001packet}
Constantinos Dovrolis, Parameswaran Ramanathan, and David Moore.
\newblock What do packet dispersion techniques measure?
\newblock In {\em INFOCOM 2001. Twentieth Annual Joint Conference of the IEEE
  Computer and Communications Societies. Proceedings. IEEE}, volume~2, pages
  905--914. IEEE, 2001.

\bibitem{dovrolis2004packet}
Constantinos Dovrolis, Parameswaran Ramanathan, and David Moore.
\newblock Packet-dispersion techniques and a capacity-estimation methodology.
\newblock {\em IEEE/ACM Transactions on Networking}, 12(6):963--977, 2004.

\bibitem{faloutsos1999power}
Michalis Faloutsos, Petros Faloutsos, and Christos Faloutsos.
\newblock On power-law relationships of the internet topology.
\newblock In {\em ACM SIGCOMM Computer Communication Review}, volume 29-4,
  pages 251--262. ACM, 1999.

\bibitem{fan2004prefix}
Jinliang Fan, Jun Xu, Mostafa~H. Ammar, and Sue~B. Moon.
\newblock Prefix-preserving ip address anonymization: Measurement-based
  security evaluation and a new cryptography-based scheme.
\newblock {\em Computer Networks}, 46(2):253--272, 2004.

\bibitem{fonseca2016measuring}
Osvaldo Fonseca, Elverton Fazzion, Italo Cunha, Pedro Henrique~Bragioni
  Las-Casas, Dorgival Guedes, Wagner Meira, Cristine Hoepers, Klaus
  Steding-Jessen, and Marcelo~H.P. Chaves.
\newblock Measuring, characterizing, and avoiding spam traffic costs.
\newblock {\em IEEE Internet Computing}, 20(4):16--24, 2016.

\bibitem{fontugne2017scaling}
Romain Fontugne, Patrice Abry, Kensuke Fukuda, Darryl Veitch, Kenjiro Cho,
  Pierre Borgnat, and Herwig Wendt.
\newblock Scaling in internet traffic: A 14 year and 3 day longitudinal study,
  with multiscale analyses and random projections.
\newblock {\em IEEE/ACM Transactions on Networking (TON)}, 25(4):2152--2165,
  2017.

\bibitem{fontugne2017pinpointing}
Romain Fontugne, Cristel Pelsser, Emile Aben, and Randy Bush.
\newblock Pinpointing delay and forwarding anomalies using large-scale
  traceroute measurements.
\newblock In {\em Proceedings of the 2017 Internet Measurement Conference},
  pages 15--28. ACM, 2017.

\bibitem{gadepally2018hyperscaling}
Vijay Gadepally, Jeremy Kepner, Lauren Milechin, William Arcand, David Bestor,
  Bill Bergeron, Chansup Byun, Matthew Hubbell, Micheal Houle, Micheal Jones,
  et~al.
\newblock Hyperscaling internet graph analysis with d4m on the mit supercloud.
\newblock {\em IEEE High Performance Extreme Computing Conference (HPEC)},
  2018.

\bibitem{heidemann2008census}
John Heidemann, Yuri Pradkin, Ramesh Govindan, Christos Papadopoulos, Genevieve
  Bartlett, and Joseph Bannister.
\newblock Census and survey of the visible internet.
\newblock In {\em Proceedings of the 8th ACM SIGCOMM Conference on Internet
  Measurement}, pages 169--182. ACM, 2008.

\bibitem{hilbert2011world}
Martin Hilbert and Priscila L{\'o}pez.
\newblock The world's technological capacity to store, communicate, and compute
  information.
\newblock {\em Science}, 332:60, 2011. doi:10.1126/science.1200970.


\bibitem{holme2015modern}
Petter Holme.
\newblock Modern temporal network theory: A colloquium.
\newblock {\em The European Physical Journal B}, 88(9):234, 2015.

\bibitem{husak2018assessing}
Martin Hus\'{a}k, Nataliia Neshenko, Morteza~Safaei Pour, Elias Bou-Harb, and
  Pavel \v{C}eleda.
\newblock Assessing internet-wide cyber situational awareness of critical
  sectors.
\newblock In {\em Proceedings of the 13th International Conference on
  Availability, Reliability and Security}, ARES 2018, pages 29:1--29:6, New
  York, 2018. ACM.

\bibitem{javed2015measurement}
Mobin Javed, Cormac Herley, Marcus Peinado, and Vern Paxson.
\newblock Measurement and analysis of traffic exchange services.
\newblock In {\em Proceedings of the 2015 Internet Measurement Conference},
  pages 1--12. ACM, 2015.

\bibitem{karagiannis2004p2p}
Thomas Karagiannis, Andre Broido, Nevil Brownlee, Kimberly~C. Claffy, and
  Michalis Faloutsos.
\newblock Is p2p dying or just hiding?
\newblock In {\em IEEE Global Telecommunications Conference, 2004,
  GLOBECOM'04.}, volume~3, pages 1532--1538. IEEE, 2004.

\bibitem{karagiannis2004transport}
Thomas Karagiannis, Andre Broido, Michalis Faloutsos, et~al.
\newblock Transport layer identification of p2p traffic.
\newblock In {\em Proceedings of the 4th ACM SIGCOMM Conference on Internet
  Measurement}, pages 121--134. ACM, 2004.

\bibitem{karvanen2003measuring}
Juha Karvanen and Andrzej Cichocki.
\newblock Measuring sparseness of noisy signals.
\newblock In {\em 4th International Symposium on Independent Component Analysis
  and Blind Signal Separation}, pages 125--130, 2003.

\bibitem{Kepner2009}
Jeremy Kepner.
\newblock {\em Parallel MATLAB for Multicore and Multinode Computers}.
\newblock SIAM, 2009.

\bibitem{kepner2011graph}
Jeremy Kepner and John Gilbert.
\newblock {\em Graph Algorithms in the Language of Linear Algebra}.
\newblock SIAM, Philadelphia, PA, 2011.

\bibitem{kepner2018mathematics}
Jeremy Kepner and Hayden Jananthan.
\newblock {\em Mathematics of Big Data: Spreadsheets, Databases, Matrices, and
  Graphs}.
\newblock MIT Press, Cambridge, MA, 2018.

\bibitem{kim2008internet}
Hyunchul Kim, Kimberly~C. Claffy, Marina Fomenkov, Dhiman Barman, Michalis
  Faloutsos, and KiYoung Lee.
\newblock Internet traffic classification demystified: Myths, caveats, and the
  best practices.
\newblock In {\em Proceedings of the 2008 ACM CoNEXT Conference}, page~11. ACM,
  2008.

\bibitem{kitsak2015long}
Maksim Kitsak, Ahmed Elmokashfi, Shlomo Havlin, and Dmitri Krioukov.
\newblock Long-range correlations and memory in the dynamics of internet
  interdomain routing.
\newblock {\em PloS one}, 10(11):e0141481, 2015.

\bibitem{kitsak2010identification}
Maksim Kitsak, Lazaros~K. Gallos, Shlomo Havlin, Fredrik Liljeros, Lev Muchnik,
  H.~Eugene Stanley, and Hern{\'a}n~A Makse.
\newblock Identification of influential spreaders in complex networks.
\newblock {\em Nature Physics}, 6(11):888, 2010.

\bibitem{kohno2005remote}
Tadayoshi Kohno, Andre Broido, and Kimberly~C. Claffy.
\newblock Remote physical device fingerprinting.
\newblock {\em IEEE Transactions on Dependable and Secure Computing},
  2(2):93--108, 2005.

\bibitem{kolda2009tensor}
Tamara~G. Kolda and Brett~W. Bader.
\newblock Tensor decompositions and applications.
\newblock {\em SIAM Review}, 51(3):455--500, 2009.

\bibitem{koliba2018governance}
Christopher~J. Koliba, Jack~W. Meek, Asim Zia, and Russell~W. Mills.
\newblock {\em Governance Networks in Public Administration and Public Policy}.
\newblock Routledge, Abingdon, 2018.

\bibitem{krioukov2012network}
Dmitri Krioukov, Maksim Kitsak, Robert~S. Sinkovits, David Rideout, David Meyer,
  and Mari{\'a}n Bogu{\~n}{\'a}.
\newblock Network cosmology.
\newblock {\em Scientific Reports}, 2:793, 2012.

\bibitem{krioukov2010hyperbolic}
Dmitri Krioukov, Fragkiskos Papadopoulos, Maksim Kitsak, Amin Vahdat, and
  Mari{\'a}n Bogun{\'a}.
\newblock Hyperbolic geometry of complex networks.
\newblock {\em Physical Review E}, 82(3):036106, 2010.

\bibitem{krioukov2009curvature}
Dmitri Krioukov, Fragkiskos Papadopoulos, Amin Vahdat, and Mari{\'a}n
  Bogu{\~n}{\'a}.
\newblock Curvature and temperature of complex networks.
\newblock {\em Physical Review E}, 80(3):035101, 2009.

\bibitem{labovitz2011internet}
Craig Labovitz, Scott Iekel-Johnson, Danny McPherson, Jon Oberheide, and Farnam
  Jahanian.
\newblock Internet inter-domain traffic.
\newblock {\em ACM SIGCOMM Computer Communication Review}, 41(4):75--86, 2011.

\bibitem{leland1994self}
Will~E. Leland, Murad~S. Taqqu, Walter Willinger, and Daniel~V. Wilson.
\newblock On the self-similar nature of ethernet traffic (extended version).
\newblock {\em IEEE/ACM Transactions on Networking (ToN)}, 2(1):1--15, 1994.

\bibitem{leskovec2005graphs}
Jure Leskovec, Jon Kleinberg, and Christos Faloutsos.
\newblock Graphs over time: densification laws, shrinking diameters and
  possible explanations.
\newblock In {\em Proceedings of the Eleventh ACM SIGKDD International
  Conference on Knowledge Discovery in Data Mining}, pages 177--187. ACM, 2005.

\bibitem{li2017fundamental}
Aming Li, Sean~P. Cornelius, Y.-Y. Liu, Long Wang, and A.-L. Barab{\'a}si.
\newblock The fundamental advantages of temporal networks.
\newblock {\em Science}, 358(6366):1042--1046, 2017.

\bibitem{li2013survey}
Bingdong Li, Jeff Springer, George Bebis, and Mehmet~Hadi Gunes.
\newblock A survey of network flow applications.
\newblock {\em Journal of Network and Computer Applications}, 36(2):567--581,
  2013.

\bibitem{li2004first}
Lun Li, David Alderson, Walter Willinger, and John Doyle.
\newblock A first-principles approach to understanding the internet's
  router-level topology.
\newblock In {\em ACM SIGCOMM Computer Communication Review}, volume~34, pages
  3--14. ACM, 2004.

\bibitem{lindmark2018minimum}
Gustav Lindmark and Claudio Altafini.
\newblock Minimum energy control for complex networks.
\newblock {\em Scientific Reports}, 8(1):3188, 2018.

\bibitem{lischke2016analyzing}
Matthias Lischke and Benjamin Fabian.
\newblock Analyzing the bitcoin network: The first four years.
\newblock {\em Future Internet}, 8(1):7, 2016.

\bibitem{liu2010tcam}
Alex~X. Liu, Chad~R. Meiners, and Eric Torng.
\newblock Tcam razor: A systematic approach towards minimizing packet
  classifiers in tcams.
\newblock {\em IEEE/ACM Transactions on Networking (TON)}, 18(2):490--500,
  2010.

\bibitem{liu2016packet}
Alex~X. Liu, Chad~R. Meiners, and Eric Torng.
\newblock Packet classification using binary content addressable memory.
\newblock {\em IEEE/ACM Transactions on Networking}, 24(3):1295--1307, 2016.

\bibitem{liu2016control}
Yang-Yu Liu and Albert-L{\'a}szl{\'o} Barab{\'a}si.
\newblock Control principles of complex systems.
\newblock {\em Reviews of Modern Physics}, 88(3):035006, 2016.

\bibitem{lu2016vital}
Linyuan L{\"u}, Duanbing Chen, Xiao-Long Ren, Qian-Ming Zhang, Yi-Cheng Zhang,
  and Tao Zhou.
\newblock Vital nodes identification in complex networks.
\newblock {\em Physics Reports}, 650:1--63, 2016.

\bibitem{lumsdaine2007challenges}
Andrew Lumsdaine, Douglas Gregor, Bruce Hendrickson, and Jonathan Berry.
\newblock Challenges in parallel graph processing.
\newblock {\em Parallel Processing Letters}, 17(01):5--20, 2007.

\bibitem{mahanti2013tale}
Aniket Mahanti, Niklas Carlsson, Anirban Mahanti, Martin Arlitt, and Carey
  Williamson.
\newblock A tale of the tails: Power-laws in internet measurements.
\newblock {\em IEEE Network}, 27(1):59--64, 2013.

\bibitem{mandelbrot1953informational}
Benoit Mandelbrot.
\newblock An informational theory of the statistical structure of language.
\newblock {\em Communication Theory}, 84:486--502, 1953.

\bibitem{medina2000origin}
Alberto Medina, Ibrahim Matta, and John Byers.
\newblock On the origin of power laws in internet topologies.
\newblock {\em ACM SIGCOMM Computer Communication review}, 30(2):18--28, 2000.

\bibitem{mirkovic2017you}
Jelena Mirkovic, Genevieve Bartlett, John Heidemann, Hao Shi, and Xiyue Deng.
\newblock Do you see me now? sparsity in passive observations of address
  liveness.
\newblock In {\em Network Traffic Measurement and Analysis Conference (TMA),
  2017}, pages 1--9. IEEE, 2017.

\bibitem{CAIDAstats}
http://www.caida.org/data/passive/trace\_stats/.

\bibitem{CAIDApubs}
http://www.caida.org/data/publications/.

\bibitem{montemurro2001beyond}
Marcelo~A. Montemurro.
\newblock Beyond the zipf--mandelbrot law in quantitative linguistics.
\newblock {\em Physica A: Statistical Mechanics and its Applications},
  300(3-4):567--578, 2001.

\bibitem{moore2003inside}
David Moore, Vern Paxson, Stefan Savage, Colleen Shannon, Stuart Staniford, and
  Nicholas Weaver.
\newblock Inside the slammer worm.
\newblock {\em IEEE Security \& Privacy}, 99(4):33--39, 2003.

\bibitem{moore2006inferring}
David Moore, Colleen Shannon, Douglas~J. Brown, Geoffrey~M. Voelker, and Stefan
  Savage.
\newblock Inferring internet denial-of-service activity.
\newblock {\em ACM Transactions on Computer Systems (TOCS)}, 24(2):115--139,
  2006.

\bibitem{moore2002code}
David Moore, Colleen Shannon, et~al.
\newblock Code-red: A case study on the spread and victims of an internet worm.
\newblock In {\em Proceedings of the 2nd ACM SIGCOMM Workshop on Internet
  measurment}, pages 273--284. ACM, 2002.

\bibitem{moore2003internet}
David Moore, Colleen Shannon, Geoffrey~M. Voelker, and Stefan Savage.
\newblock Internet quarantine: Requirements for containing self-propagating
  code.
\newblock In {\em INFOCOM 2003. Twenty-Second Annual Joint Conference of the
  IEEE Computer and Communications. IEEE Societies}, volume~3, pages
  1901--1910. IEEE, 2003.

\bibitem{mucha2010community}
Peter~J. Mucha, Thomas Richardson, Kevin Macon, Mason~A. Porter, and Jukka-Pekka
  Onnela.
\newblock Community structure in time-dependent, multiscale, and multiplex
  networks.
\newblock {\em Science}, 328(5980):876--878, 2010.

\bibitem{newman2001clustering}
Mark~E.J. Newman.
\newblock Clustering and preferential attachment in growing networks.
\newblock {\em Physical Review E}, 64(2):025102, 2001.

\bibitem{olston2010web}
Christopher Olston, Marc Najork, et~al.
\newblock Web crawling.
\newblock {\em Foundations and Trends{\textregistered} in Information
  Retrieval}, 4(3):175--246, 2010.

\bibitem{paluch2018fast}
Robert Paluch, Xiaoyan Lu, Krzysztof Suchecki, Boles{\l}aw~K. Szyma{\'n}ski, and
  Janusz~A Ho{\l}yst.
\newblock Fast and accurate detection of spread source in large complex
  networks.
\newblock {\em Scientific Reports}, 8(1):2508, 2018.

\bibitem{papadopoulos2012popularity}
Fragkiskos Papadopoulos, Maksim Kitsak, M.~{\'A}ngeles Serrano, Mari{\'a}n
  Bogun{\'a}, and Dmitri Krioukov.
\newblock Popularity versus similarity in growing networks.
\newblock {\em Nature}, 489(7417):537, 2012.

\bibitem{paxson1996end}
Vern Paxson.
\newblock End-to-end routing behavior in the internet.
\newblock {\em ACM SIGCOMM Computer Communication Review}, 26(4):25--38, 1996.

\bibitem{paxson1995wide}
Vern Paxson and Sally Floyd.
\newblock Wide area traffic: the failure of Poisson modeling.
\newblock {\em IEEE/ACM Transactions on Networking (ToN)}, 3(3):226--244, 1995.

\bibitem{perc2017statistical}
Matja{\v{z}} Perc, Jillian~J. Jordan, David~G. Rand, Zhen Wang, Stefano
  Boccaletti, and Attila Szolnoki.
\newblock Statistical physics of human cooperation.
\newblock {\em Physics Reports}, 687:1--51, 2017.

\bibitem{prasad2003bandwidth}
Ravi Prasad, Constantinos Dovrolis, Margaret Murray, and Kimberly~C.~Claffy.
\newblock Bandwidth estimation: metrics, measurement techniques, and tools.
\newblock {\em IEEE Network}, 17(6):27--35, 2003.

\bibitem{rabinovich2016measuring}
Michael Rabinovich and Mark Allman.
\newblock Measuring the internet.
\newblock {\em IEEE Internet Computing}, 20(4):6--8, 2016.

\bibitem{reuther2018interactive}
Albert Reuther, Jeremy Kepner, Chansup Byun, Siddharth Samsi, William Arcand,
  David Bestor, Bill Bergeron, Vijay Gadepally, Michael Houle, Matthew Hubbell,
  et~al.
\newblock Interactive supercomputing on 40,000 cores for machine learning and
  data analysis.
\newblock {\em IEEE High Performance Extreme Computing Conference (HPEC)},
  2018.

\bibitem{saito2000sparsity}
Naoki Saito, Brons~M. Larson, and Bertrand B{\'e}nichou.
\newblock Sparsity vs. statistical independence from a best-basis viewpoint.
\newblock In {\em Wavelet Applications in Signal and Image Processing VIII},
  volume 4119, pages 474--487. International Society for Optics and Photonics,
  2000.

\bibitem{saleh2006modeling}
Osama Saleh and Mohamed Hefeeda.
\newblock Modeling and caching of peer-to-peer traffic.
\newblock In {\em Proceedings of the 2006
  14th IEEE International Conference on Network Protocols, ICNP'06}, pages 249--258. IEEE, 2006.

\bibitem{schaeffer2007graph}
Satu~Elisa Schaeffer.
\newblock Graph clustering.
\newblock {\em Computer Science Review}, 1(1):27--64, 2007.

\bibitem{sheridan2018preferential}
Paul Sheridan and Taku Onodera.
\newblock A preferential attachment paradox: How preferential attachment
  combines with growth to produce networks with log-normal in-degree
  distributions.
\newblock {\em Scientific Reports}, 8(1):2811, 2018.

\bibitem{soule2004identify}
Augustin Soule, Antonio Nucci, Rene Cruz, Emilio Leonardi, and Nina Taft.
\newblock How to identify and estimate the largest traffic matrix elements in a
  dynamic environment.
\newblock In {\em ACM SIGMETRICS Performance Evaluation Review}, volume~32,
  pages 73--84. ACM, 2004.

\bibitem{spring2002measuring}
Neil Spring, Ratul Mahajan, and David Wetherall.
\newblock Measuring isp topologies with rocketfuel.
\newblock {\em ACM SIGCOMM Computer Communication Review}, 32(4):133--145,
  2002.

\bibitem{stumpf2012critical}
Michael~P.H. Stumpf and Mason~A. Porter.
\newblock Critical truths about power laws.
\newblock {\em Science}, 335(6069):665--666, 2012.

\bibitem{tune2013internet}
Paul Tune, Matthew Roughan, H.~Haddadi, and O.~Bonaventure.
\newblock Internet traffic matrices: A primer.
\newblock {\em Recent Advances in Networking}, 1:1--56, 2013.

\bibitem{virkar2014power}
Yogesh Virkar and Aaron Clauset.
\newblock Power-law distributions in binned empirical data.
\newblock {\em The Annals of Applied Statistics}, pages 89--119, 2014.

\bibitem{wang2016statistical}
Zhen Wang, Chris~T. Bauch, Samit Bhattacharyya, Alberto d'Onofrio, Piero
  Manfredi, Matja{\v{z}} Perc, Nicola Perra, Marcel Salath{\'e}, and Dawei
  Zhao.
\newblock Statistical physics of vaccination.
\newblock {\em Physics Reports}, 664:1--113, 2016.

\bibitem{willinger2009mathematics}
Walter Willinger, David Alderson, and John~C. Doyle.
\newblock Mathematics and the internet: A source of enormous confusion and
  great potential.
\newblock {\em Notices of the American Mathematical Society}, 56(5):586--599,
  2009.

\bibitem{willinger2002scaling}
Walter Willinger, Ramesh Govindan, Sugih Jamin, Vern Paxson, and Scott Shenker.
\newblock Scaling phenomena in the internet: Critically examining criticality.
\newblock {\em Proceedings of the National Academy of Sciences}, 99(suppl
  1):2573--2580, 2002.

\bibitem{willinger1997self}
Walter Willinger, Murad~S. Taqqu, Robert Sherman, and Daniel~V. Wilson.
\newblock Self-similarity through high-variability: Statistical analysis of
  ethernet lan traffic at the source level.
\newblock {\em IEEE/ACM Transactions on Networking (ToN)}, 5(1):71--86, 1997.

\bibitem{xu2012}
Zongben Xu, Xiangyu Chang, Fengmin Xu, and Hai Zhang.
\newblock $ l\_ $\{$1/2$\}$ $ regularization: A thresholding representation
  theory and a fast solver.
\newblock {\em IEEE Transactions on Neural Networks and Learning Systems},
  23(7):1013--1027, 2012.

\bibitem{Rahimi2018scale}
Hongbo~Dong Yaghoub~Rahimi, Chao~Wang and Yifei Lous.
\newblock A scale invariant approach for sparse signal recovery.
\newblock {\em arXiv preprint arXiv:1812.08852}, 2018.

\bibitem{yu2012predicted}
Shui Yu, Guofeng Zhao, Wanchun Dou, and Simon James.
\newblock Predicted packet padding for anonymous web browsing against traffic
  analysis attacks.
\newblock {\em IEEE Transactions on Information Forensics and Security},
  7(4):1381--1393, 2012.

\bibitem{yu2017link}
Xuecheng Yu and Tianguang Chu.
\newblock Link prediction from partial observation in scale-free networks.
\newblock In {\em Chinese Intelligent Systems Conference}, pages 199--205.
  Springer, 2017.

\bibitem{zhang2014named}
Lixia Zhang, Alexander Afanasyev, Jeffrey Burke, Van Jacobson, Patrick Crowley,
  Christos Papadopoulos, Lan Wang, Beichuan Zhang, et~al.
\newblock Named data networking.
\newblock {\em ACM SIGCOMM Computer Communication Review}, 44(3):66--73, 2014.

\bibitem{zhang2005estimating}
Yin Zhang, Matthew Roughan, Carsten Lund, and David~L. Donoho.
\newblock Estimating point-to-point and point-to-multipoint traffic matrices:
  An information-theoretic approach.
\newblock {\em IEEE/ACM Transactions on Networking (TON)}, 13(5):947--960,
  2005.

\end{thebibliography}
